# SoK: All You Ever Wanted to Know About x86/x64 Binary Disassembly But Were Afraid to Ask

Chengbin Pang[*][‡][§]  Ruotong Yu[*]  Yaohui Chen[†]  Eric Koskinen[*]  Georgios Portokalidis[*]  Bing Mao[‡]  Jun Xu[*]

[*]Stevens Institute of Technology  [†]Facebook Inc.  [‡]Nanjing University

*Abstract*—Disassembly of binary code is hard, but necessary for improving the security of binary software. Over the past few decades, research in binary disassembly has produced many tools and frameworks, which have been made available to researchers and security professionals. These tools employ a variety of strategies that grant them different characteristics. The lack of systematization, however, impedes new research in the area and makes selecting the right tool hard, as we do not understand the strengths and weaknesses of existing tools. In this paper, we systematize binary disassembly through the study of nine popular, open-source tools. We couple the manual examination of their code bases with the most comprehensive experimental evaluation (thus far) using 3,788 binaries. Our study yields a comprehensive description and organization of strategies for disassembly, classifying them as either *algorithm* or else *heuristic*. Meanwhile, we measure and report the impact of individual algorithms on the results of each tool. We find that while principled algorithms are used by all tools, they still heavily rely on heuristics to increase code coverage. Depending on the heuristics used, different coverage-vs-correctness trade-offs come in play, leading to tools with different strengths and weaknesses. We envision that these findings will help users pick the right tool and assist researchers in improving binary disassembly.

## I. INTRODUCTION

The disassembly of binary programs is a crucial task in reverse engineering and software security, and it is a core component of innumerable works on malware analysis [54], code-similarity measurement [17, 42, 55], vulnerability discovery [25, 66, 82, 95], security retrofitting [2, 78, 80, 100, 104, 109, 111] and patching [11]. However, correctly disassembling a binary is *challenging*, mainly owing to the loss of information (*e.g.,* symbols and types) occurring when compiling a program to machine code and the complexity of constructs (*e.g.,* jump tables, data embedded in code, *etc.*) used to efficiently implement language features.

Binary disassembly has seen remarkable advancements in the past decade, awarding researchers and developers with a variety of tools and frameworks, under both open source [3, 33, 90, 94, 95, 102, 103] and commercial [36, 74] licenses. These tools have lifted a significant burden off researchers that aim to develop new, advanced binary analysis techniques. This new plurality of options encapsulates a broad variety of underlying strategies with different guarantees, which fall under two categories:

[§]Pang is a PhD student at Nanjing University. This work was done while Pang was a Visiting Scholar at Stevens Institute of Technology.

TABLE I: The group of open-source tools that our study covers and representative works that use those tools.

| Tool (Version) | Source (Release Date) | Public Use |
|---|---|---|
| PSI (1.0) | Website [63] (Sep 2014) | [50, 88, 111] |
| UROBOROS (0.11) | Github [93] (Nov. 2016) | [103] |
| DYNINST (9.3.2) | Github [79] (April 2017) | [7, 18, 69, 73, 96] |
| OBJDUMP (2.30) | GNU [47] (Jan. 2018) | [21, 103, 111] |
| GHIDRA (9.0.4) | Github [75] (May 2019) | [24, 45, 91] |
| MCSEMA (2.0.0) | Github [13] (Jun. 2019) | [22, 41, 44] |
| ANGR (8.19.7.25) | Github [8] (Oct. 2019) | [20, 71, 81, 98, 112] |
| BAP (2.1.0) | Github [26] (Mar. 2020) | [10, 16, 64] |
| RADARE2 (4.4.0) | Github [89] (April 2020) | [4, 31, 52, 58] |

- **Algorithms** typically produce results with some correctness guarantees. They mostly leverage knowledge from the binary (*e.g.,* symbols), the machine (*e.g.,* instruction set), and/or the ABI (*e.g.,* calling conventions).
- **Heuristics** are based on common patterns and typically do not offer assurances of correctness.

Moreover, each tool adopts a different set of strategies, with technical details not always fully documented or publicized. To complicate the matters, the implemented strategies have evolved over time, further deviating from documentation. The above have created a knowledge gap that impedes the users of these tools and, specifically, binary analysis researchers. To bridge the gap, we must answer several questions:

- *Q1 – What are the algorithms and heuristics used in existing disassembly tools and how do they interact?*
- *Q2 – What is the coverage & accuracy of heuristic methods in comparison to algorithmic ones? Are there trade-offs?*
- *Q3 – What errors do existing disassembly tools make and what are the underlying causes?*

To answer these questions, this paper presents a systematization of binary disassembly research, through the study of nine popular open-source tools shown in Table I. Unlike past research [5, 56, 68, 77, 105], we study these tools both qualitatively and quantitatively to understand the tools not only as a whole, but also their individual algorithms and heuristics.

More specifically, our qualitative study of the tools is based on manually inspecting source code. This allows us to answer *Q1* by presenting their exact and most recent strategies, avoiding ambiguities and out-of-date information found in documentation and publications. The quantitative study answers questions *Q2–Q3* by applying the tools on a corpus of 3,788 benchmark binaries, consisting of utilities, client/server programs, and popular libraries on both Linux and Windows systems (see Table IV). To evaluate the tools in terms of coverage and accuracy, we built an analysis framework based on LLVM, GCC, the Gold Linker, and Visual Studio to automatically collect the ground truth while building the corpus. We evaluate the tools by individually measuring different disassembly phases to quantify the effectiveness of the strategies employed. Our evaluation presents the degree of use, precision, and pitfalls of each component of each tool.

By systematically dissecting and evaluating the tools, we were able to make new observations that amend or complement prior knowledge. Our major observations include: (1) For better coverage, mainstream tools incorporate heuristics in nearly every phase of disassembly. These heuristics are heavily used in disassembling real-world binaries and, without them, the tools cannot provide practical utility in many tasks. (2) Heuristics typically cannot provide correctness assurances and lead to various errors, particularly when encountering complex constructs. Moreover, previous works may have overestimated the reliability of those heuristics. For instance, a recent study [5] (unintentionally) overstated the accuracy of linear sweep because many benchmarks containing data-in-code were not considered. (3) Tools may share the same group of algorithms and heuristics, however, they organize and combine them differently, leading to different accuracy-coverage trade-offs. (4) Tools have different strengths across different tasks. For instance, commercial tools are better at recovering instructions but open source tools can better identify cross-references.

**Contributions:** Our main contributions are as follows:
- We present a thorough systematization of binary disassembly from the perspective of algorithms and heuristics. To our knowledge, this is the first research that can answer *Q1–Q3*.
- We developed a compiler-based framework for automated end-to-end collection of ground truth for binary disassembly. We used it to compose a benchmark data set for assessing binary disassembly tools. The framework and benchmarks are available at https://github.com/junxzm1990/x86-sok.
- We present, to our knowledge, the most comprehensive evaluation of open-source disassembly tools. Our analysis unveils the prevalence of heuristics, their contribution to disassembly, and shortcomings.
- We make new observations and improve the understanding of binary-disassembly strategies and tools. We envision that these insights will facilitate future research in the area of disassembly and drive improvements in existing tools.

## II. SCOPE OF SYSTEMATIZATION

### A. Functionality

In general, binary disassembly can involve different tasks based on the context of use. This work focuses on tasks that relate to binary-software security. Table II classifies popular security works and summarizes the information each class needs to recover from binary code. Our study accordingly concentrates on the disassembly aspects of providing them:

**Disassembly** is the process of recovering the assembly instructions of a binary. Perfect disassembly separates data from code regions and correctly identifies the instructions that were emitted by the compiler or introduced by the developer.

TABLE II: Popular solutions of binary security and the information the solutions need from binary disassembly. "Inst", "CFG", "Func", and "Xrefs" respectively mean legitimate instructions, control flow graph, functions, and cross references.

| Category | Solutions & Required Information | |
|---|---|---|
| Vulnerability Finding | [25, 34, 97] | Inst, CFG, Func, Xrefs |
| Control Flow Integrity | [37, 80, 100, 108, 109, 111] [19, 38, 51, 72, 85, 87, 99] | Inst, CFG, Func |
| Code Layout Randomization | [28, 53, 60, 104] [61, 65, 78, 106, 109] | Inst, CFG, Func, Xrefs |
| Execute-only Code | [21, 110] | Inst, CFG |
| Legacy-code Patching | [11, 101, 102, 103] | Inst, CFG, Func, Xrefs |
| Code Similarity Measurement | [14, 17, 40, 42, 55, 57, 82] [15, 29, 30, 43, 62, 70, 83] | Inst, CFG, Func |
| Software Fault Isolation | [38, 39, 67, 107] | Inst, CFG, Func |
| Software De-bloating | [46, 86, 92] | Inst, CFG, Func |

**Symbolization** determines cross-references (*xrefs* for short) or precisely, numeric values in the binary that are references of other code or data objects. Depending on the location of the reference and the location of the target, there are four types of xrefs: code-to-code (c2c), code-to-data (c2d), data-to-code (d2c), and data-to-data (d2d).

**Function Entry Identification** locates the entry points of functions. A special but important case is the *main* function.

**CFG Reconstruction** re-builds the control flow graph (CFG) of a binary program. We consider direct control transfers, indirect jumps/calls, tail calls, and non-returning functions.

### B. Targeted Binaries

Similarly to the majority of the works we study, we focus on binaries with the following key properties: (1) They have been produced with mainstream compilers and linkers; (2) Binaries may include hand-written assembly; (3) They have not been obfuscated; (4) We do not assume symbol availability, *i.e.,* binaries are stripped; (5) We only consider X86/X64 binaries. The majority of effort in prior works has focused on such binaries, owing to the popularity of the architectures; (6) They run on Linux or Windows operating systems.

### C. Targeted Tools

Our systematization is based on study of disassembly tools. We use five criteria to select tools: (1) They are designated for disassembly or have an independent module for disassembly. (2) They can do automated disassembly without user interactions. (3) They are open source tools so that we can study their implemented strategies. (4) They have unique strategies that are not fully covered by other tools. (5) They can run our targeted binaries to support our quantitative evaluation.

Following the above criteria, we selected 9 tools, as listed in Table I. We also looked at JakStab [59], RetDec [32], and BinCat [12] but excluded them in the study because (1) JakStab cannot run our benchmark binaries due to a parsing error (https://github.com/jkinder/jakstab/issues/9); (2) RetDec aims for de-compilation; it uses preliminary strategies for disassembly, which are all covered by other tools we selected; (3) BinCat requires user interactions to do disassembly. The disassembly results vary upon different user interactions.

TABLE III: The specifics of existing algorithms (numbered with rings like ①) and heuristics (numbered with discs like ❶).

| Alg. | | Algorithms & Heuristics | Goals | Tools |
|---|---|---|---|---|
| Disassembly | Linear Sweep | ① Start from code addresses with symbols | Code accuracy | OBJDUMP, PSI, UROBOROS |
| | | ❶ Continuous scanning for instructions | Code coverage | OBJDUMP, PSI, UROBOROS |
| | | ❷ Skip bad opcodes | Code accuracy | OBJDUMP |
| | | ❸ Replace padding and re-disassembly | Code accuracy | PSI |
| | | ❹ Exclude code around errors | Code accuracy | UROBOROS |
| | Recursive Descent | ② Follow control flow to do disassembly | Code accuracy | DYNINST, GHIDRA, ANGR, BAP, RADARE2 |
| | | ③ Start from program entry, main, and symbols | Code accuracy | DYNINST, GHIDRA, ANGR, BAP, RADARE2 |
| | | ❺ Function entry matching | Code coverage | DYNINST, GHIDRA, ANGR, BAP, RADARE2 |
| | | ❻ Linear sweep code gaps | Code coverage | ANGR |
| | | ❼ Disassembly from targets of xrefs | Code coverage | GHIDRA, RADARE2 |
| Symbolization | Xrefs | ④ Exclude data units that are floating points | Xref accuracy | ANGR |
| | | ❽ Brute force operands and data units | Xref coverage | UROBOROS, MCSEMA, GHIDRA, ANGR |
| | | ❾ Pointers in data have machine size | Xref accuracy | UROBOROS, MCSEMA, GHIDRA, ANGR |
| | | ❿ Alignment of pointers in data | Xref accuracy | UROBOROS, MCSEMA, GHIDRA |
| | | ⓫ Pointers in data or referenced by other xrefs can be non-aligned | Xref coverage | GHIDRA, ANGR |
| | | ⓬ References to code can only point to function entries | Xref accuracy | GHIDRA |
| | | ⓭ Enlarge boundaries of data regions | Xref coverage | GHIDRA, ANGR |
| | | ⓮ Address tables have minimal size of 2 | Xref accuracy | GHIDRA |
| | | ⓯ Exclude pointers that may overlap with a string | Xref accuracy | MCSEMA, GHIDRA |
| | | ⓰ While scanning data regions, use step-length based on type inference | Xref accuracy | ANGR |
| Function Entry | MAIN Function | ⑤ Identify main based on arguments to __libc_start_main | Func coverage | ANGR, BAP |
| | | ⓱ Identify main using patterns in _start/__scrt_common_main_seh | Func coverage | DYNINST, RADARE2 |
| | General Function | ⑥ Identify function entries based on symbols | Func coverage | DYNINST, GHIDRA, ANGR, BAP, RADARE2 |
| | | ⑦ Identify function entries based on exception information | Func coverage | GHIDRA |
| | | ⑧ Identify function entries based on targets of direct calls | Func coverage | DYNINST, GHIDRA, ANGR, BAP, RADARE2 |
| | | ⑨ Identify function entries by resolving indirect calls | Func coverage | GHIDRA, ANGR |
| | | ⓲ Identify function entries based on prologues/decision-tree | Func coverage | DYNINST, GHIDRA, ANGR, BAP, RADARE2 |
| | | ⓳ Consider begins of code discovered by linear scan as function entries | Func coverage | ANGR |
| CFG | Indirect Jump | ⑩ Use VSA to resolve jump table targets | CFG accuracy | DYNINST, GHIDRA, ANGR |
| | | ⓴ Follow patterns to determine jump tables | CFG accuracy | DYNINST, GHIDRA, RADARE2 |
| | | ㉑ Discard jump tables with index bound larger than a threshold | CFG accuracy | GHIDRA, ANGR, RADARE2 |
| | | ㉒ Restrict the depth of slice for VSA | Efficiency | DYNINST, ANGR |
| | Indirect Call | ⑪ Identify targets based on constant propagation | CFG coverage | GHIDRA, ANGR |
| | Tail Call | ⑫ Consider a jump to the start of another function as a tail call | CFG accuracy | DYNINST, ANGR |
| | | ㉓ Determine tail call based on distance between the jump and its target | CFG accuracy | RADARE2 |
| | | ㉔ A tail call and its target cross multiple functions | CFG accuracy | GHIDRA |
| | | ㉕ Tail calls cannot be conditional jumps | CFG accuracy | GHIDRA, ANGR |
| | | ㉖ A tail call tears down its stack | CFG accuracy | DYNINST, ANGR |
| | | ㉗ A tail call does not jump to the middle of a function | CFG accuracy | ANGR |
| | | ㉘ Target of a tail call cannot be target of any conditional jumps | CFG accuracy | ANGR |
| | Non-returning Function | ⑬ Identify system calls or library functions that are known non-returning | CFG accuracy | DYNINST, GHIDRA, ANGR, BAP, RADARE2 |
| | | ⑭ Identify functions with no ret and no tail calls that return | CFG accuracy | DYNINST, ANGR, RADARE2 |
| | | ⑮ Identify functions that always call non-returning functions | CFG accuracy | BAP |
| | | ㉙ Detect non-returning functions based on fall-through after the call-sites | CFG accuracy | GHIDRA |

## III. ANALYSIS OF TOOLS

To understand the strategies employed in today's binary disassembly tools, we studied 9 representative examples in Table I. These tools have varying popularity and cover nearly all publicly known techniques in binary disassembly [105].

Our investigation is primarily based on studying source code, instead of solely relying on available documentation and publications. Source code reflects the exact semantics of the strategies applied, protecting us from ambiguities in the documents. Also, many tools have evolved over time and their corresponding documentation and publications are out-of-date.

The rest of this section presents our findings, summarized in Table III. We assign a number to each *algorithm* and *heuristic*, respectively placing it within a ring (*e.g.,* ①) or disc (*e.g.,* ❶).

### A. Algorithms and Heuristics in Disassembly

The disassembly strategies we study fall into two broad and well-known classes: linear sweep and recursive descent.

**Linear Sweep** [OBJDUMP, PSI, UROBOROS]: Linear sweep continuously scans pre-selected code ranges and identifies valid instructions (❶), exploiting the rationale that modern assemblers tend to layout code successively to reduce the binary's size. In general, a linear sweep strategy can be described by *how it selects sweep ranges* and *how it handles errors during scanning*. As such, we summarize algorithms according to these two aspects.

All tools in this class follow OBJDUMP to select code regions for sweep: they process code ranges specified by symbols in the .symtab and .dynsym sections (①), followed by remaining gaps in the code sections. In a general sense, these ranges comprehensively encapsulate legitimate instructions.

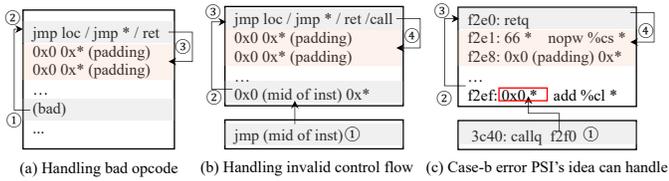

(a) Handling bad opcode  (b) Handling invalid control flow  (c) Case-b error PSI's idea can handle

Fig. 1: Error-handling by PSI. Part (a) and part (b) respectively show the handling of bad opcode and invalid control transfers. Part (c) is an invalid control-transfer from BinUtils that PSI should handle but the actual implementation does not.

Various heuristics are used for error detection and handling. OBJDUMP deems invalid opcodes as errors, skips a byte, and resumes scanning (❷). Beyond invalid opcodes, PSI considers control transfers to non-instructions as errors. Furthermore, PSI has more sophisticated error-handling as shown in Fig. 1. Upon a bad opcode, PSI traces backwards from the erroneous instruction to a non-fall-through control transfer (unconditional jump, indirect jump, or return) and identifies padding after that control transfer. Replacing the padding with nop instructions, PSI then re-runs linear-sweep for re-disassembly (❸). When encountering an invalid control transfer, the public version of PSI only handles cases where the transfer part is correct but code around the target contains errors. Specifically, it checks the code around the target, seeks a preceding instruction that starts with zero, and finally identifies a further preceding non-fall-through control transfer and its following padding. Again, PSI will replace the padding with nop instructions and re-run linear-sweep. UROBOROS follows a similar idea as PSI. But instead of re-disassembly, UROBOROS simply excludes the code around the error locations (❹).

The design of PSI can handle cases like the one in Fig. 1(c). Its implementation, however, is too restrictive for high effectiveness. For instance, the public version of PSI only considers zero-started padding and cannot correct the error in Fig. 1(c).

> To sum up, linear sweep aggressively scans all possible code and hence, maximizes the recovery of instructions. However, it can run into errors due to data-in-code. To address errors, existing tools rely on heuristics for correction, which are less comprehensive and can have limited utility (§ IV-B1).

**Recursive Descent** [DYNINST, GHIDRA, ANGR, BAP, RADARE2]: Recursive descent starts with a given code address and performs disassembly following the control flow (②). Strategies in this category usually consist of three components: (1) how to select code addresses, (2) how to resolve control flow, and (3) how to handle the code gaps left by recursive disassembly. Accordingly, we summarize the existing tools based on the three components.

All the tools we study consider the program entry and available symbols as code addresses for recursive disassembly (③). These addresses are, in principle, known to be safe. Further, ANGR, BAP, DYNINST and RADARE2 also discover the main function and the details are covered in § III-C.

When encountering direct control transfers, the tools expand the disassembly to the targets. However, to handle indirect control transfers, different tools adopt different approaches. We will cover the details in § III-D. Another challenging part of control flow is to determine non-returning functions. Related details are also discussed in § III-D.

As indirect control flows are (formally and practically) undecidable, recursive descent often leaves behind code gaps. Our evaluation in § IV-B1 shows that recursive disassembly alone can miss **49.35%** of the code on average. As such, existing tools incorporate heuristics to enlarge code coverage, which inevitably jeopardizes correctness guarantees. The most common heuristic, used by ANGR, DYNINST, RADARE2, BAP, and GHIDRA, searches for function entry points in the code gaps based on common function prologues / epilogues or pre-trained decision-tree models [9] (❺). On finding a function entry, the tools will consider it as a new starting point for recursive disassembly. According to our evaluation (§ IV-B3), function-entry matching on average identifies **17.36%** of all the functions, leading to a **31.55%** increase of code coverage.

Beyond function matching, existing tools also use heuristics that are more aggressive. ANGR performs linear sweep on the code gaps and recursive disassembly on legitimate instructions (❻). In case of errors, ANGR skips the current basic block and moves to the next bytes. This linear scan increases the code coverage of ANGR by around **8.20%**. However, it will misidentify data as code (*e.g.,* Listing 2 in Appendix G). GHIDRA includes targets of xrefs in recursive descent (❼). This strategy discovers **4.33%** more code coverage. However, the xrefs are mostly collected with heuristics, which can lead to errors like Listing 3 in Appendix G.

> In summary, strict recursive descent ensures correctness but often produces insufficient coverage. To expand code coverage, existing tools incorporate many aggressive heuristics that undermine the correctness guarantees (§ IV-B1).

*B. Algorithms and Heuristics in Symbolization*

Symbolization identifies numerical values in the binary that are actually references to code or data objects. Tools generally follow the workflow in Fig. 2.

**Constant Operand and Data Unit Extraction** [ANGR, GHIDRA, UROBOROS, MCSEMA]: These tools start by identifying numerical values that are potential pointers. They search through all instructions to identify constant operands and scan the non-code regions to find data units (❽). As identification of constant operands is trivial, we omit the details and focus on data units. In general, a data unit is composed of consecutive $n$-bytes at an aligned address. However, different tools have varying choices of $n$, alignment, and non-code regions:

- All tools assume that a data unit's size is the same as the machine size, *i.e.,* 4 bytes in 32-bit and 8 bytes in 64-bit binaries (❾). This assumption, however, is not always true. Listing 4 in Appendix G shows a jump table with 4-byte entries from a 64-bit binary. In such cases, the assumption about a data unit's size can mislead symbolization in tools that do not otherwise handle jump tables (*e.g.,* UROBOROS).

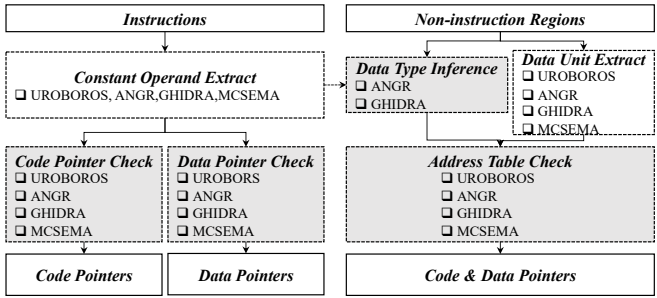

Fig. 2: A general workflow of symbolization.

- UROBOROS and MCSEMA use machine-size alignment (❿). GHIDRA assumes a 4-byte alignment unless a data unit is the target of another xref. In the latter case, GHIDRA has no alignment requirement (⓫). ANGR enforces no alignment requirement (⓫) due to observations of unaligned pointers [101]. Our evaluation shows that the choice of alignment is a coverage-accuracy trade-off: around 600 pointers are saved at non-aligned addresses while the no alignment assumption leads to nearly 60% of ANGR's false positives.
- Besides data segments, GHIDRA and ANGR also search for data units from non-disassembled code regions.

**Data Unit Type Inference** [ANGR, GHIDRA]: In operand extraction, ANGR and GHIDRA infer the types of data units when possible. ANGR identifies memory loads from data units. If the loaded values flow to floating-point instructions, ANGR marks the data units as floating points. This inference is in general reliable as it follows data flow. GHIDRA uses a more aggressive strategy: given a constant operand "pointing" to a data unit, GHIDRA considers the data unit to be the start of a string if it is followed by a sequence of ASCII/Unicode bytes and a null-byte. Otherwise, GHIDRA deems the data unit to be a pointer if the value inside meets the following conditions: (*1*) the value is at least `4096`; (*2*) the value is an address of an instruction or an address in a non-code region; (*3*) if the value is an address of an instruction in a known function, the instruction must be the function entry (⓬); (*4*) if the value is a data address, the address cannot overlap another typed data unit. GHIDRA's type inference has no correctness assurance.

**Code-to-Code and Code-to-Data Xrefs** [ANGR, GHIDRA, UROBOROS, MCSEMA]: For each constant operand, ANGR, UROBOROS, and MCSEMA seek to symbolize it as a code pointer, checking whether the operand refers to a legitimate instruction. Beyond, GHIDRA has two extra rules: (*1*) the operand cannot be a value in {[0-4095], 0xffff, 0xff00, 0xffffff, 0xff0000, 0xff0000, 0xffffffff, 0xffffff00, 0xffff0000, 0xff00000}; (*2*) the instruction being referred to must be a function entry (if the function was known) (⓬). We measured heuristic ⓬ with 3,788 binaries. We discover thousands of pointers to the middle of functions (*e.g.,* pointers for try-catch in exception handling), showing that heuristic ⓬ is unsound.

For a constant operand that cannot be a code pointer, the tools attempt to symbolize it as a data pointer, checking if the operand refers to a legitimate data location. In the checking process, ANGR enlarges the boundaries of a data region by 1,024 bytes and GHIDRA adopts a similar idea (⓭) because many pointers are dereferenced with an offset. This strategy indeed benefits coverage (*e.g.,* Listing 6 in Appendix G). It, however, also introduces errors like Listing 5 in Appendix G. Recall that GHIDRA leverages xrefs to aid recursive descent (§ III-A). When a constant operand is symbolized as a xref to a non-disassembled code region, GHIDRA recursively disassembles that region. If GHIDRA runs into errors like bad opcodes and invalid control transfers, it rolls back the disassembly.

**Address Table** [ANGR, GHIDRA, UROBOROS, MCSEMA]: Beyond constant operands, these tools also symbolize the non-code regions by locating *address tables*: a group of consecutive data units that are pointers. In general, determining address tables depends on the choice of table size and the rules to classify a data unit as a pointer. GHIDRA considers 2 as the minimal size of an address table (⓮) and the others consider 1. While the choice of GHIDRA helps more accurately identify grouped pointers like function tables, it misses many individual pointers, leading to false negatives. With regard to determining pointers, all tools follow the approaches as we previously discussed. After the initial generation of address tables, ANGR, MCSEMA and GHIDRA also apply refinements:

- ANGR excludes table entries that are floating points (④).
- MCSEMA excludes table entries that may overlap with a string (⓯). When a piece of data can be both a pointer and a string, MCSEMA prefers string. As we will discuss below, ANGR uses an opposite strategy.
- GHIDRA excludes table entries that point to the middle of recovered functions. GHIDRA also excludes table entries that overlap with strings or cut into other pointers. Finally, GHIDRA splits an address table when adjacent entries have a distance larger than 0xffffff.
- Given an entry to non-disassembled code, GHIDRA expands the recursive descent using the aforementioned approach.

ANGR uses a special strategy when brute-force searching data regions. Given a location, ANGR *in turn* checks whether the data inside is a pointer, a ASCII/Unicode string, or an arithmetic sequence. If any type matches, ANGR jumps over the typed bytes and then resumes the search (⓰). This strategy incurs many false negatives like Listing 7 in Appendix G.

> Overall, there is a lack of algorithmic solutions to symbolization. Today's tools incorporate a multitude of heuristics, striving for a coverage-correctness balance.

### C. Algorithms and Heuristics in Function Entry Identification

Most tools use separate strategies to identify the entry of `main` versus the entries of other functions. As such, we first discuss `main`, followed by the other types of functions.

**Main Function** [DYNINST, ANGR, BAP, RADARE2]: To locate it, ANGR and BAP analyze the `_start` function and,

```
1 48 c7 c7 e2 e0 40 00   mov $0x40e0e2,%rdi ;main
2 ff 15 ce 48 05 00  **  callq __libc_start_main
```

Listing 1: Call to `__libc_start_main` in `_start`.

following calling conventions, infer the first argument passed by `_start` to `__libc_start_main` (⑤). Take Listing 1

as an example. ANGR and BAP locate the instruction at line 1 and deem the immediate operand (`0x40e0e2`) to be the address of `main`. As `__libc_start_main` is a standard interface, ANGR and BAP ensure correctness. RADARE2 and DYNINST search architecture-specific patterns near the call to `__libc_start_main` to get the address of `main` (⑰). DYNINST finds the instruction right before the call and extracts the immediate operand; RADARE2 searches the address after a fixed sequence of raw bytes (*e.g.,* `48 c7 c7` in Fig. 1). For Windows binaries, RADARE2 seeks a pattern in `__scrt_common_main_seh` as it does for Linux binaries to locate `main`. Other tools do not particularly find `main`.

**General Functions** [DYNINST, GHIDRA, ANGR, BAP, RADARE2]: To identify the entries of non-`main` functions, these tools adopt a hybrid approach that consists of three parts:

*(1)* The tools seek symbols remaining in the `.symtab` and `.dynsym` sections to determine known-to-be-good functions (⑥). (Only) GHIDRA considers the `.eh_frame` section to identify functions that have unwinding information (⑦). As mandated by X86-64 ABI, modern compilers (*e.g.,* LLVM and GCC) keep unwinding information for every function. As we illustrate in § IV-B3, this way GHIDRA can identify nearly all function entries, but (surprisingly) it does not use exception information when handling Windows binaries.

*(2)* All tools consider targets of direct calls to be function entries (⑧), while ANGR and GHIDRA additionally resolve certain indirect calls to determine more function entries (⑨). Finally, DYNINST, GHIDRA, ANGR and RADARE2 include targets of tail calls as function entries (see § III-D).

*(3)* All tools use pattern-based approaches to further recover functions (⑱). GHIDRA, ANGR and RADARE2 find function entries based on common prologues (or epilogues); DYNINST (by default) and BAP find function entries with pre-trained decision-tree models [9]. As we will present in § IV-B3, this pattern-based approach is heavily used and indeed identifies many functions. However, it is very sensitive to architectures, optimizations, and compiler specifics.

ANGR adopts an extra aggressive approach: during linear scan over code gaps left by recursive descent, it treats the begin of each identified code piece as a new entry (⑲). This method improves the coverage but incurs many errors (§ IV-B3).

> To sum up, the identification of function entry mostly uses hybrid approaches, mixing algorithms and heuristics.

### D. Algorithms and Heuristics in CFG Reconstruction

CFG reconstruction consists of many tasks. We focus on the challenging ones: resolving indirect jumps/calls, detecting tail calls, and finding non-returning functions. Direct jumps/calls are not discussed as they are easily derivable after disassembly.

**Indirect Jumps** [DYNINST, GHIDRA, ANGR, RADARE2]: From our benchmark binaries (Table IV), we observed three types of indirect jumps: (1) jump tables [23] (compiled from `switch-case` and `if-else` statements); (2) indirect tail calls (indirect calls optimized as tail calls); and (3) hand-written ones (*e.g.,* `longjmp` and other cases in Glibc [48]).

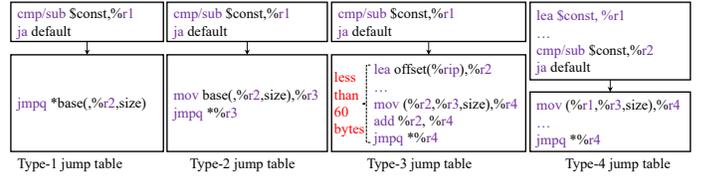

Fig. 3: Handling of jump tables by RADARE2.

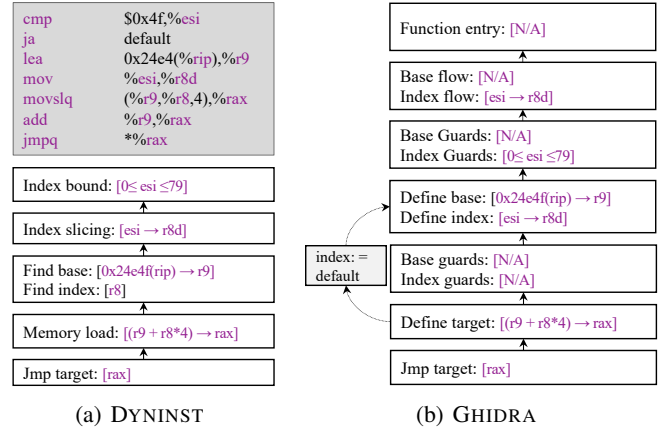

Fig. 4: Jump table resolution in DYNINST and GHIDRA. The *upper* half of Sub-Fig. 4a shows a jump table from Gold Linker. *Lower* half of Sub-Fig. 4a and Sub-Fig. 4b respectively illustrate how DYNINST and GHIDRA resolve the jump table.

RADARE2 only handles jump tables, by searching four types of patterns shown in Fig. 3 (⑳). Consider Type-1 jump table as an example: encountering a jump table of this type, RADARE2 searches an indirect jump in the format of `jmp [base + reg * size]` and a `cmp`/`sub` instruction in the preceding basic block. On finding the two items, RADARE2 deems `base` as the base address and the constant operand in `cmp`/`sub` as the upper bound of the index. If that upper bound exceeds 512, RADARE2 discards the jump table (㉑). The strategies of RADARE2 are highly sensitive to compiling configurations and are less effective with resolving jump tables (§ IV-B4).

DYNINST also only handles jump tables, using a hybrid approach shown in Fig. 4a. It performs backward slicing from the target. In the sliced area, if the first memory read (after simplification) has the format of `[CONST + reg * size]` (⑳), DYNINST deems the indirect jump to be a jump table, respectively using `CONST` and `reg` as the base address and index. Starting from the index, DYNINST performs backwards slicing again up to 50 assignments [35] (㉒) or the function entry. In this slice, DYNINST uses a simplified Value Set Analysis (VSA) to gather value bounds along the flow of the index (and its aliases) (⑩).

GHIDRA first considers an indirect jump to be a jump table and resolves it with a strategy shown in Fig. 4b. In the current function, GHIDRA seeks a single path that defines both the base address and the index (⑳). Along the path, GHIDRA tracks the propagation of the base and index to identify their value bounds. Instead of using a full VSA, GHIDRA considers restrictions by variable types, conditional jumps, and `and` instructions. Similar to RADARE2, GHIDRA discards jump tables with an index bound over 1024 (㉑). If GHIDRA cannot resolve the jump table, it will consider the indirect jump as an indirect call and perform analysis that we will discuss shortly.

ANGR, given an indirect jump, considers the operand as a source and runs backwards slicing. In the sliced area, ANGR uses full-scale VSA to identify possible targets (⑩). However, the public version of ANGR restricts the slicing to at most three levels of basic blocks (㉒), trading utility for efficiency. ANGR also adopts heuristic ㉑ with a very large threshold: 100,000.

> In summary, tools employ various heuristics to resolve indirect jumps. These heuristics, mostly derived for accuracy, introduce fewer errors but have limited coverage (§ IV-B4).

**Indirect Calls** [GHIDRA, ANGR]: GHIDRA finds targets of indirect calls based on constant propagation (⑪). It tracks the intra-procedure propagation of constants from immediate operands, the LEA instructions, and global memory. Once a constant flows to an indirect call, GHIDRA takes the constant as a target. ANGR also uses constant propagation (⑪) to handle indirect calls but only considers the current basic block.

**Tail Calls** [RADARE2, ANGR, DYNINST, GHIDRA]: For efficiency, function calls at the end of procedures are often optimized as jumps (*i.e., tail calls*). Tools adopt different strategies to detect tail calls.

RADARE2 uses a simple heuristic to determine tail calls: the distance between a jump and its target exceeding a certain threshold (㉓). This heuristic exploits the insight that different functions are usually apart. However, it is hard to pick a threshold that is both effective and accurate.

GHIDRA determines a jump as a tail call if the code between the jump and its target spans multiple functions (㉔). The heuristic can lead to both false positives due to discontinuous functions and false negatives due to unrecognized functions. GHIDRA further excludes conditional jumps from consideration (㉕), preventing the detection of 21.6% of the tail calls.

DYNINST takes a sophisticated strategy. It considers a jump as a tail call if the target is the start of a known function (⑫). Otherwise, DYNINST checks two rules to determine a tail call: (*1*) the jump's target cannot be reached by only following false branches; (*2*) right before the jump, the stack is tore down by [leave; pop $reg] or [add $rsp $const] (㉖). While the first rule is hard to reason, the second rule exploits an intrinsic property of tail calls: the current function recycles the stack such that the child re-uses its return address. The above pattern-based approach is less accurate, leading to 97% of DYNINST's false positives and most of its false negatives.

ANGR adopts similar strategies as DYNINST. It first identifies jumps whose targets are starts of known functions (⑫). Beyond that, ANGR requires four conditions to detect a tail call: (*1*) the jump is unconditional (㉕); (*2*) the stack at the jump is tore down based on stack height analysis (㉖); (*3*) the target does not belong to any function or it belongs to the current function (㉗); (*4*) all incoming edges of the target are unconditional jumps or direct calls (㉘). As we will show in § IV-B4, these heuristics do not provide correctness guarantees and each can lead to both false positives and false negatives.

> Tools take different strategies to detect tail calls. These strategies depend critically on function entry detection, inheriting inaccuracies from function identification.

**Non-returning Functions** [ANGR, BAP, RADARE2, DYNINST, GHIDRA]: Tools use a similar workflow for detecting non-returning functions. First, they gather the group of library functions or system calls that are known to be non-returning (⑬). Second, from this initial group of functions, tools further find other ones.

ANGR, RADARE2, and DYNINST use the same idea: they scan each function and, if no ret instruction is found, they consider the function to be non-returning (⑭). However, they adopt different solutions when encountering a call to a child function that has unknown returning status: (1) ANGR simply assumes a fall-through after the call; (2) RADARE2 takes a similar strategy but when the child status is later updated, RADARE2 recursively propagates the update to its predecessor functions; (3) DYNINST takes a depth-first approach to first handle the successor functions. Only until the status of the child function is determined, DYNINST will continue handling the original call. To our understanding, the strategies in RADARE2 and DYNINST are equivalent. Both are principled and produce nearly perfect precision (§ IV-B4).

GHIDRA also follows the idea of (⑭). To handle a call with unknown returning status, GHIDRA checks the fall-through code after the call. If the code runs into abnormal regions (*i.e.,* data, unrecognizable instructions, or another function) or the code is referenced by xrefs (other than jumps), GHIDRA marks the fall-through unsafe. If the child function of that call incurs at least three unsafe fall-through, GHIDRA considers the child to be non-returning. Further, instead of doing recursive updates, GHIDRA simply runs the detection process twice.

BAP considers a function to be non-returning if all the paths in the function end at a call to a non-returning function (⑭). To handle calls to child functions with unknown status, it takes a similar recursive-updating strategy as RADARE2.

> Tools use many principled strategies to detect non-returning functions, ensuring higher correctness.

## IV. LARGE-SCALE EVALUATION

### A. Evaluation Setup

**Benchmarks:** We used the software listed in Table IV to experimentally evaluate the tools we studied. The software includes programs and libraries of diverse functionality and complexity, written in C / C++, and containing hand-written assembly and hard-coded binary code. It also carries a significant amount of complex constructs as listed in Table XVII, fitting the use as a benchmark. To test the effect of different compilers and options, we built each software package for two operating systems (Linux and Windows), using three compilers (GCC-8.1.0 and LLVM-6.0.0 on Linux, and Visual Studio 2015 on Windows), various optimization levels (O0, O2, O3, Os, Ofast on Linux, Od, O1, O2, Ox on Windows), and two architectures (X86 and X64). This resulted in 3,788 binaries, which are less than the theoretically possible 4484 (177×20 + 59×16), as some cannot be built for certain configurations.
**Ground Truth:** We intercept the compiling and linking process to automatically obtain ground truth about the produced binaries, which includes instructions, functions, control transfers, jump tables, xrefs, and remaining complex constructs.

TABLE IV: Software used for evaluating tools.

| Type | Name | Programs/Binaries | |
|---|---|---|---|
| | | Linux | Windows |
| Benchmark | SPEC CPU2006 | 30 / 546 | 15 / 120 |
| Utilities | Unzip-6.0 Coreutils-8.30 7-zip-19 Findutils-4.4 Binutils-2.26 Tiff-4.0 | 125 / 2500 | 26 / 196 |
| Clients | Openssl-1.1.0l Putty-0.73 D8-6.4 Filezilla-3.44.2 Busybox-1.31 Protobuf-c-1 ZSH-5.7.1 VIM-8.1 XML2-2.9.8 Openssh-8.0 Git-2.23 | 13 / 154 | 13 / 104 |
| Servers | Lighttpd-1.4.54 Mysqld-5.7.27 Nginx-1.15.0 SQLite-3.32.0 | 3 / 49 | 2 / 16 |
| Libraries | Glibc-2.27 libpcap-1.9.0 libv8-6.4 libtiff-4.0.10 libxml2-2.9.8 libsqlite-3.32.0 libprotobuf-c-1.3.2 | 6 / 79 | 3 / 24 |
| Total | | 177 / 3328 | 59 / 460 |

On Linux, we replicate the approach in CCR [60]. CCR extends the LLVM Machine Code layer to record the needed information in each compiled bitcode/assembly file. It also instruments the GNU Gold Linker to merge the information during linking. Throughout experimentation, we identified several limitation of the CCR linker, which we addressed by extending it to: (1) collect information from previously ignored sections (.text.unlikely, .text.exit,.text.startup,.text.hot); (2) support relocatable objects (compiled with -r); (3) record both the size and the offset of basic blocks; (4) handle linker-inserted functions (*e.g.,* _start) and statically-linked Glibc functions; (5) support X86 targets. We also extend the CCR approach to GCC by instrumenting its RTL pass to label all related constructs when producing assembly code. We also customized the GNU Assembler to record the locations of the constructs and xrefs in the emitted object files. Finally, we re-use the CCR linker to merge object files.

On Windows, we combine compiler options, symbol/debug information, and lightweight manual analysis to build ground truth. Details are covered in Appendix B.
**Tools and Configurations:** Besides open-source tools, we also test two commercial tools, IDA PRO-7.4 and BINARY NINJA-1.2. We detail the configuration of all tools in Appendix C. It should be noted that we run two versions of GHIDRA and ANGR, including a version of GHIDRA not using exception information, namely ***GHIDRA-NE***, and a version of ANGR without linear scan, namely ***ANGR-NS***.

## B. Evaluation Results & Analysis

*1) Disassembly :* This evaluation measures the recovery of legitimate instructions. We excluded all the padding bytes and linker-inserted functions (*e.g.,* _start). We also inserted a symbol of main so that all recursive tools can find it.

TABLE V: Evaluation results of instruction recovery. **L** and **W** are short for Linux and Windows. `Pre` and `Rec` means precision and recall. `Ave`/`Min` show the average/minimal results among all binaries. The best/worst results specific to each optimization level are respectively marked as blue/red.

| | | Instructions | | | | | Instructions | | | |
|---|---|---|---|---|---|---|---|---|---|---|
| | L | Avg | | Min | | W | Avg | | Min | |
| | | Pre | Rec | Pre | Rec | | Pre | Rec | Pre | Rec |
| Objdump | O0 | 99.98 | 99.99 | 88.80 | 99.70 | Od | 98.04 | 99.98 | 86.84 | 99.83 |
| | O2 | 99.94 | 99.94 | 84.76 | 99.56 | O1 | 98.96 | 99.97 | 83.84 | 99.84 |
| | O3 | 99.95 | 99.95 | 85.84 | 93.45 | O2 | 97.58 | 99.97 | 83.73 | 99.82 |
| | Os | 99.95 | 99.97 | 84.30 | 96.45 | Ox | 97.57 | 99.97 | 83.73 | 99.82 |
| | Of | 99.90 | 99.90 | 86.52 | 96.65 | - | - | - | - | - |
| Dyninst | O0 | 99.99 | 99.07 | 99.90 | 6.30 | - | - | - | - | - |
| | O2 | 99.99 | 92.45 | 99.46 | 4.15 | - | - | - | - | - |
| | O3 | 99.99 | 91.72 | 99.51 | 19.47 | - | - | - | - | - |
| | Os | 99.99 | 89.79 | 99.83 | 5.26 | - | - | - | - | - |
| | Of | 99.99 | 91.65 | 99.07 | 19.55 | - | - | - | - | - |
| McSema | O0 | 99.99 | 99.99 | 99.96 | 99.80 | Od | 99.79 | 97.49 | 94.82 | 83.20 |
| | O2 | 99.99 | 99.97 | 99.88 | 99.11 | O1 | 99.83 | 99.32 | 95.06 | 97.47 |
| | O3 | 99.99 | 99.83 | 99.90 | 98.64 | O2 | 99.84 | 95.41 | 95.81 | 75.07 |
| | Os | 99.99 | 99.99 | 99.84 | 99.35 | Ox | 99.82 | 96.16 | 94.18 | 78.30 |
| | Of | 99.99 | 99.83 | 99.85 | 96.49 | - | - | - | - | - |
| Ghidra-NE | O0 | 99.99 | 98.03 | 99.60 | 63.26 | Od | 99.88 | 96.01 | 94.84 | 65.05 |
| | O2 | 99.99 | 80.65 | 99.64 | 2.06 | O1 | 99.86 | 90.40 | 95.04 | 73.67 |
| | O3 | 99.99 | 77.83 | 99.69 | 17.77 | O2 | 99.88 | 92.12 | 95.80 | 72.00 |
| | Os | 99.99 | 89.66 | 99.76 | 12.76 | Ox | 99.87 | 92.09 | 94.10 | 69.34 |
| | Of | 99.99 | 78.48 | 99.70 | 17.92 | - | - | - | - | - |
| Ghidra | O0 | 99.99 | 99.65 | 99.96 | 64.51 | Od | 99.88 | 96.01 | 94.84 | 65.05 |
| | O2 | 99.99 | 94.90 | 99.68 | 51.31 | O1 | 99.86 | 90.40 | 95.04 | 73.67 |
| | O3 | 99.99 | 93.63 | 99.72 | 48.26 | O2 | 99.88 | 92.12 | 95.80 | 72.00 |
| | Os | 99.99 | 95.26 | 99.73 | 16.23 | Ox | 99.87 | 92.09 | 94.10 | 69.34 |
| | Of | 99.99 | 93.73 | 99.73 | 37.78 | - | - | - | - | - |
| Angr-NS | O0 | 99.99 | 98.79 | 99.96 | 64.88 | Od | 99.97 | 93.02 | 98.67 | 14.50 |
| | O2 | 99.99 | 87.75 | 99.89 | 48.75 | O1 | 99.97 | 93.42 | 98.10 | 18.94 |
| | O3 | 99.99 | 89.00 | 99.90 | 54.31 | O2 | 99.96 | 90.31 | 98.14 | 17.80 |
| | Os | 99.99 | 94.58 | 99.94 | 60.73 | Ox | 99.96 | 90.17 | 98.14 | 17.80 |
| | Of | 99.99 | 88.59 | 99.87 | 51.07 | - | - | - | - | - |
| Angr | O0 | 99.99 | 99.97 | 99.95 | 97.64 | Od | 98.96 | 99.92 | 90.01 | 99.37 |
| | O2 | 99.99 | 99.98 | 99.79 | 99.23 | O1 | 99.33 | 99.90 | 87.71 | 99.16 |
| | O3 | 99.99 | 99.98 | 99.81 | 99.30 | O2 | 98.66 | 99.88 | 87.55 | 99.20 |
| | Os | 99.99 | 99.99 | 99.77 | 99.78 | Ox | 98.64 | 99.88 | 87.55 | 99.20 |
| | Of | 99.98 | 99.98 | 99.84 | 99.75 | - | - | - | - | - |
| Bap | O0 | 99.99 | 82.50 | 99.96 | 9.37 | Od | 99.98 | 71.44 | 98.97 | 30.24 |
| | O2 | 99.98 | 66.88 | 99.70 | 9.81 | O1 | 99.96 | 78.95 | 98.54 | 50.87 |
| | O3 | 99.99 | 65.64 | 99.69 | 11.73 | O2 | 99.90 | 70.49 | 97.48 | 42.41 |
| | Os | 99.98 | 76.92 | 97.20 | 9.16 | Ox | 99.90 | 70.47 | 97.45 | 38.75 |
| | Of | 99.94 | 66.62 | 95.56 | 12.64 | - | - | - | - | - |
| Radare2 | O0 | 99.98 | 87.92 | 94.51 | 44.59 | Od | 99.21 | 86.25 | 46.59 | 30.00 |
| | O2 | 99.99 | 78.80 | 99.29 | 8.55 | O1 | 99.82 | 78.60 | 92.87 | 46.21 |
| | O3 | 99.98 | 75.32 | 99.24 | 9.23 | O2 | 98.50 | 75.78 | 42.78 | 44.73 |
| | Os | 99.97 | 85.31 | 91.09 | 8.59 | Ox | 98.48 | 75.60 | 45.74 | 38.44 |
| | Of | 99.99 | 76.50 | 99.24 | 8.90 | - | - | - | - | - |
| IDA | O0 | 99.99 | 99.99 | 99.61 | 99.34 | Od | 99.80 | 99.98 | 94.83 | 99.54 |
| | O2 | 99.99 | 99.95 | 99.76 | 98.15 | O1 | 99.82 | 99.96 | 95.08 | 99.39 |
| | O3 | 99.99 | 99.95 | 99.87 | 98.03 | O2 | 99.84 | 99.89 | 95.83 | 96.44 |
| | Os | 99.99 | 94.61 | 99.93 | 57.31 | Ox | 99.82 | 99.90 | 94.20 | 96.66 |
| | Of | 99.99 | 93.04 | 99.88 | 51.62 | - | - | - | - | - |
| Ninja | O0 | 99.99 | 99.81 | 99.72 | 89.85 | Od | 99.61 | 99.23 | 94.84 | 96.61 |
| | O2 | 99.99 | 98.79 | 99.50 | 72.01 | O1 | 99.76 | 99.74 | 95.14 | 96.04 |
| | O3 | 99.98 | 97.38 | 99.56 | 71.77 | O2 | 99.51 | 98.96 | 95.93 | 93.26 |
| | Os | 99.99 | 99.46 | 99.37 | 77.51 | Ox | 99.52 | 99.22 | 94.32 | 96.94 |
| | Of | 99.98 | 97.49 | 99.65 | 71.79 | - | - | - | - | - |

**Overall Performance:** In Table V, we show the overall results of instruction recovery. Note that we do not show the results of PSI and UROBOROS as they are close to OBJDUMP.

In general, the results of disassembly vary across tool categories. Linear tools and linear-sweep aided tools, like OBJDUMP and ANGR, have every high coverage (99.95%+ recall). Recursive tools have lower recovery rates and some can only recover less than 80% of the instructions (BAP and RADARE2). We also notice that the performance of recursive tools changes across optimization levels and architectures. In particular, nearly all recursive tools (ANGR-NS, GHIDRA-NE, DYNINST, BAP, RADARE2) have reduced recovery as the optimization increases and when analyzing X64 targets. This is because that optimization levels and architectures affect the function matching in recursive tools, further leading to missing of instructions. Such results well comply with previous observations [5].

In the aspect of precision, we have an opposite observation. Recursive tools have high precision (over 99.5%), regardless of the compiler, architecture, and optimization level. Linear tools are less precise. The precision of OBJDUMP, in the worst case, drops to around 85%. This difference is mainly because recursive tools mostly follow the control flow, ensuring correctness. However, linear tools scan every byte and often run into errors when data appears in code. For instance, OBJDUMP produces the worst result in analyzing `Openssl` (precision: 85.35%) because `Openssl` has a lot of data in assembly files and OBJDUMP wrongly identifies the data as code.

**Use of Heuristics:** To augment the correctness of linear sweep, PSI introduces error detection and handling. In our evaluation, PSI can analyze 971 of the X86 binaries on Linux systems. On the binaries, the linear sweep produces over 10K errors. PSI captures 26% of the errors leading to bad opcode and other 6% resulting in invalid control transfers. However, the public version of PSI has strict requirement of padding patterns, preventing it from fixing the errors. In summary, PSI's heuristic can capture 32% of the errors in linear sweep. However, we are unable to report the detectable errors due to PSI's implementation restrictions.

In contrast, heuristics in recursive tools are mostly for coverage enhancement. We measure the effectiveness of each heuristic in turn. We start with pure recursive descent by disabling function matching, linear scan in ANGR, xref aided disassembly in GHIDRA and RADARE2. The results are shown in Table XVIII in Appendix D. Unsurprisingly, all tools have nearly perfect precision. However, without heuristics, the tools have very low recall. ANGR, GHIDRA, DYNINST all have a recall around 51% and RADARE2 recovers no more than 10% of the code. We noted that GHIDRA still produces high recall with Linux binaries. This is because GHIDRA uses exception information to highly accurately identify functions (§ III-C).

On top of pure recursive descent, we in turn enable function matching, linear scan, and the use of xrefs. With function matching, recursive tools recover significantly more functions and code (see Table XIX in Appendix D). In particular, ANGR-NS and DYNINST respectively identify 21.30% and 18.24% more functions, leading to recovery of 36.44% and 26.65% more code (comparing Table V and XVIII). Second, according to the results of ANGR and ANGR-NS in Table V, linear scan aids ANGR to recover 8.20% more instructions. Third, the use of xrefs in GHIDRA leads to a 4.33% increase in code coverage, as shown by the result of GHIDRA on Linux binaries in Table V and XVIII.

TABLE VI: Statistics of false positives in instruction recovery. *Pad*, *data*, *Func*, *Non-Ret*, *Jump-Tbl* respectively represent errors due to padding, data-in-code, wrong function matching, non-returning functions, and wrong jump tables.

| Tools | Percentage of False Positives (%) | | | | | |
|---|---|---|---|---|---|---|
| | Pad | Data | Func | Non-Ret | Jump-Tbl | Other |
| **Objdump** | 71.7 | 28.3 | 0.0 | 0.0 | 0.0 | 0.0 |
| **Dyninst** | 0.0 | 0.0 | 36.2 | 39.9 | 23.8 | 0.0 |
| **Ghidra** | 0.0 | 0.0 | 18.2 | 65.4 | 3.0 | 13.4 |
| **Ghidra-NE** | 0.0 | 0.0 | 6.4 | 61.7 | 5.9 | 26.0 |
| **Angr** | 10.5 | 10.3 | 76.3 | 2.9 | 0.0 | 0.0 |
| **Angr-NS** | 0.0 | 0.0 | 70.2 | 29.1 | 0.7 | 0.0 |
| **Bap** | 0.0 | 0.0 | 89.0 | 0.8 | 0.0 | 10.2 |
| **Radare2** | 0.0 | 0.0 | 96.5 | 2.2 | 1.3 | 0.0 |

**Understanding of Errors:** In Table VI, we summarize the statistic of false positives. For linear tools (*e.g.,* OBJDUMP), all the false positives are caused by misidentifying padding bytes or data-in-code as code. For recursive tools, the most common reasons of errors include (1) considering illegal locations as function entries; (2) missing non-returning functions and assuming the calls to them fall through; (3) incorrect resolution of jump tables. Beyond that, ANGR's linear sweep incurs 21% of the errors due to data in code; BAP and GHIDRA have a few implementation flaws, also leading to a group of errors.

The reasons of false negatives are consistent. All the false negatives by linear tools are side-effects of false positives. For recursive tools, most false negatives are caused by undetected function: as shown in Table X, recursive tools averagely miss 25.0% of the functions. The remaining instructions are missed mainly because certain jump tables are not resolved and false positives over-run the legitimate instructions.

*2) Symbolization:* In this evaluation, we measure the recovery of xrefs. Xrefs in direct jumps/calls are trivial to identify so we excluded them. We also excluded xrefs in wrong instructions since such errors are rooted from incorrect disassembly. Finally, we did not consider jump tables as they are separately measured in § IV-B4.

**Overall Performance:** We summarize the overall performance of symbolization in Table VII. In general, open-source tools have much higher recovery rates (98.35% on average) than commercial tools (88.63% on average). The difference is even bigger with the worst case. This is because open-source tools brute force all constant operands and data units while the commercial tools adopt more conservative strategies. Our hypothesis can be verified by comparing MCSEMA and IDA PRO, since MCSEMA just adds a round of brute force on top of IDA PRO.

Somewhat surprisingly, open-source tools also have high precision (99.92% on average). Based on our observation, the high precision is because (1) the heuristic-based checks are generally restrictive and (2) most benchmark programs have less data. On programs that have more data, the tools are more inclined to make mistakes (*e.g.,* `Mysqld`, having plenty of data, leads to the lowest precision in many tools).

TABLE VII: Evaluation results of symbolization. UROBOROS cannot run Linux-Of binaries and all Window binaries. We also omitted ANGR's symbolization on Windows binaries as ANGR's Reassembler component cannot run them.

| | | Symbolizations | | | | | Symbolizations | | | |
| --- | --- | --- | --- | --- | --- | --- | --- | --- | --- | --- |
| | L | Avg | | Min | | W | Avg | | Min | |
| | | Pre | Rec | Pre | Rec | | Pre | Rec | Pre | Rec |
| *Urobo | O0 | 99.99 | 100 | 99.89 | 100 | - | - | - | - | - |
| | O2 | 100 | 100 | 100 | 100 | - | - | - | - | - |
| | O3 | 99.99 | 100 | 99.91 | 100 | - | - | - | - | - |
| | Os | 99.99 | 100 | 99.88 | 100 | - | - | - | - | - |
| | Of | - | - | - | - | - | - | - | - | - |
| McSema | O0 | 99.99 | 99.77 | 99.04 | 94.05 | Od | 99.93 | 96.05 | 94.79 | 78.22 |
| | O2 | 99.99 | 99.96 | 98.78 | 97.87 | O1 | 99.96 | 96.08 | 98.00 | 84.80 |
| | O3 | 99.99 | 99.96 | 98.88 | 98.79 | O2 | 99.97 | 95.38 | 98.70 | 86.92 |
| | Os | 99.98 | 99.75 | 98.68 | 90.91 | Ox | 99.95 | 95.51 | 96.54 | 83.42 |
| | Of | 99.96 | 99.68 | 93.10 | 81.82 | - | - | - | - | - |
| Ghidra | O0 | 99.90 | 95.70 | 61.99 | 45.52 | Od | 99.70 | 96.53 | 94.75 | 78.12 |
| | O2 | 99.91 | 95.53 | 87.35 | 37.71 | O1 | 99.63 | 95.02 | 95.59 | 77.86 |
| | O3 | 99.92 | 95.99 | 88.36 | 41.60 | O2 | 99.86 | 95.25 | 98.67 | 74.54 |
| | Os | 99.89 | 95.37 | 61.98 | 41.72 | Ox | 99.83 | 95.30 | 96.45 | 74.52 |
| | Of | 99.97 | 96.26 | 98.19 | 20.00 | - | - | - | - | - |
| Angr | O0 | 99.86 | 99.52 | 81.74 | 95.38 | - | - | - | - | - |
| | O2 | 99.88 | 99.54 | 87.28 | 95.35 | - | - | - | - | - |
| | O3 | 99.90 | 99.59 | 88.29 | 96.21 | - | - | - | - | - |
| | Os | 99.81 | 99.99 | 86.57 | 98.96 | - | - | - | - | - |
| | Of | 99.84 | 99.99 | 88.05 | 98.89 | - | - | - | - | - |
| IDA | O0 | 99.98 | 95.96 | 97.17 | 36.96 | Od | 99.99 | 95.76 | 99.90 | 82.51 |
| | O2 | 99.93 | 94.32 | 86.77 | 31.96 | O1 | 99.99 | 94.11 | 99.88 | 72.65 |
| | O3 | 99.94 | 95.06 | 88.38 | 31.96 | O2 | 99.99 | 94.26 | 99.88 | 69.94 |
| | Os | 99.93 | 93.81 | 85.59 | 25.26 | Ox | 99.99 | 94.39 | 99.89 | 70.48 |
| | Of | 99.98 | 95.63 | 97.52 | 32.02 | - | - | - | - | - |
| Ninja | O0 | 99.99 | 79.89 | 99.81 | 17.29 | Od | 99.90 | 85.37 | 93.62 | 40.94 |
| | O2 | 99.99 | 81.08 | 97.59 | 19.01 | O1 | 99.98 | 84.67 | 99.61 | 45.18 |
| | O3 | 99.99 | 81.38 | 99.40 | 21.15 | O2 | 99.98 | 84.54 | 99.63 | 50.17 |
| | Os | 99.99 | 77.71 | 98.67 | 17.45 | Ox | 99.97 | 84.98 | 99.44 | 50.14 |
| | Of | 99.99 | 82.35 | 99.52 | 21.15 | - | - | - | - | - |

**Use of Heuristics:** As shown in Table III, symbolization involves many heuristics, competing for a coverage-correctness trade-off. Our evaluation shows that heuristic ❽ (the brute-force based approach), ⓫ (the no-alignment assumption about pointers), and ⓭ (the enlargement of data regions) bring full coverage of xrefs in our benchmarks. The other heuristics are instead striving for correctness. In the following, we discuss them in turn.

Heuristic ❽ (the brute-force based approach) is necessary for symbolization. This can be verified by comparing IDA PRO and MCSEMA. MCSEMA adds a round of brute force to IDA PRO, increasing the coverage from 95% to 98%. Heuristic ❾ (pointers in data have machine size) is principled if jump tables are not considered. From over 6 million xrefs, we observe no violations. Heuristic ❿ (pointers in data are aligned) and ⓫ (pointers in data may not be aligned) conflict with each other, and neither of them is perfect. Heuristic ❿ misses around 600 xrefs while heuristic ⓫ introduces most false positives in ANGR (over 50K). Heuristic ⓬ (references to code point to function entries) can reduce false positives but it misses thousands of xrefs to the middle of functions (*e.g.,* pointers for try-catch in exception handling). Heuristic ⓭ (enlarging boundaries of data regions) helps recover 12K xrefs in data, but leads to 6 and 2K+ false positives in ANGR and GHIDRA. Heuristic ⓮ introduces most false negatives in GHIDRA, although removing a small set of false positives. Heuristic ⓯ produces over 3K false negatives in MCSEMA because the inference of strings is not accurate.

TABLE VIII: Statistics of false positives in symbolization. *Align*, *Type*, *Type-based Sliding*, and *Extended-data* respectively mean no alignment (heuristic 11), ANGR's moving scheme (heuristic 16), enlarging data regions (heuristic 13).

| Tools | Percentage of False Positives (%) | | | |
| --- | --- | --- | --- | --- |
| | Align | Type-based Sliding | Extended-data | Collision |
| Urobo* | 0.00 | 0.00 | 0.00 | 100 |
| Mcsema | 0.00 | 0.00 | 0.00 | 100 |
| Ghidra | 0.29 | 0.00 | 3.91 | 95.80 |
| Angr | 58.58 | 4.47 | 0.01 | 36.94 |

TABLE IX: Statistics of false negatives in symbolization. *Align*, *Type*, *Extended-data*, *Function*, *Address-table* respectively mean assumption of alignment (heuristic 10), preferring strings over pointers (heuristic 15), enlarging data regions (heuristic 13), and size of address table (heuristic 14).

| Tools | Percentage of False Negative (%) | | | | |
| --- | --- | --- | --- | --- | --- |
| | Align | Type | Extended-data | Address Table | Func |
| Urobo* | 0.00 | 0.00 | 0.00 | 0.00 | 0.00 |
| Mcsema | 1.14 | 75.22 | 23.64 | 0.00 | 0.00 |
| Ghidra | 0.01 | 3.13 | 0.00 | 96.53 | 0.33 |
| Angr | 0.00 | 100 | 0.00 | 0.00 | 0.00 |

**Understanding of Errors:** Table VIII shows the statistics of false positives. Using the assumption that pointers in data can be non-aligned, ANGR and GHIDRA trigger 59% and 0.3% of their false positives. They also produce false positives because they enlarge the boundaries of data regions when checking the legitimacy of xrefs' targets. All the other false positives are due to collisions between numeric values and pointers.

As shown in Table IX, most false negatives by ANGR and MCSEMA are because they exclude pointers that overlap with the inferred strings. MCSEMA also misses 23.64% of the xrefs that point to locations outside of the data regions. GHIDRA, because of its assumption about the address table's minimal size and that code pointers always point to function entries, respectively produces 96.53% and 0.33% of its false negatives.

*3) Function Entry Identification:* This evaluation measures the identification of function entries. In this test, we further considered NUCLEUS [6] and BYTEWEIGHT [9]. We re-trained BYTEWEIGHT with our benchmarks binaries.

**Overall Performance:** Table X presents the overall results. The key observation is that function entry identification remains a challenge. 4 of the open source tools can only identify less than 80% of the functions. In particular, RADARE2 has a recall lower than 66%. Such results indicate, even with heuristics, we have yet to develop better function identification. We also observe that the results of function identification varies across optimization levels and architectures. This is because the tools widely use signature-based function matching, which are specific to optimizations and architectures. Besides limited coverage, existing tools also have lower precision in function identification, comparing to instruction recovery and symbolization. For instance, ANGR and BAP respectively have an average precision of 56.67% and 86.11%.

TABLE X: Evaluation results of function entry identification.

|  | L | Function Entry | | | | W | Function Entry | | | |
|---|---|---|---|---|---|---|---|---|---|---|
|  |  | Avg | | Min | | | Avg | | Min | |
|  |  | Pre | Rec | Pre | Rec |  | Pre | Rec | Pre | Rec |
| Dyninst | O0 | 99.65 | 97.11 | 82.76 | 2.07 | - | - | - | - | - |
|  | O2 | 93.92 | 75.26 | 48.95 | 12.37 | - | - | - | - | - |
|  | O3 | 93.37 | 74.12 | 48.95 | 15.95 | - | - | - | - | - |
|  | Os | 98.98 | 50.46 | 66.67 | 4.00 | - | - | - | - | - |
|  | Of | 98.70 | 42.76 | 50.00 | 6.25 | - | - | - | - | - |
| Ghidra-NE | O0 | 99.92 | 97.29 | 94.59 | 45.30 | Od | 98.68 | 83.41 | 66.42 | 40.58 |
|  | O2 | 99.40 | 48.46 | 88.89 | 10.00 | O1 | 98.45 | 69.78 | 61.41 | 40.62 |
|  | O3 | 99.30 | 47.52 | 88.89 | 10.26 | O2 | 98.48 | 70.00 | 61.55 | 40.59 |
|  | Os | 98.85 | 69.97 | 77.78 | 10.50 | Ox | 98.43 | 70.93 | 55.53 | 40.32 |
|  | Of | 98.32 | 49.23 | 81.82 | 11.02 | - | - | - | - | - |
| Ghidra | O0 | 99.94 | 99.66 | 86.82 | 55.11 | Od | 98.68 | 83.41 | 66.42 | 40.58 |
|  | O2 | 97.25 | 87.92 | 71.92 | 16.94 | O1 | 98.45 | 69.78 | 61.41 | 40.62 |
|  | O3 | 96.48 | 87.59 | 71.30 | 20.00 | O2 | 98.48 | 70.00 | 61.55 | 40.59 |
|  | Os | 99.22 | 88.90 | 90.00 | 12.60 | Ox | 98.43 | 70.93 | 55.53 | 40.32 |
|  | Of | 95.90 | 88.31 | 71.27 | 20.69 | - | - | - | - | - |
| Angr-NS | O0 | 97.47 | 96.92 | 45.65 | 60.00 | Od | 79.19 | 78.18 | 27.07 | 10.63 |
|  | O2 | 86.37 | 68.08 | 44.74 | 17.01 | O1 | 77.86 | 69.36 | 27.04 | 10.90 |
|  | O3 | 85.23 | 68.35 | 42.15 | 17.00 | O2 | 76.04 | 66.60 | 24.03 | 10.71 |
|  | Os | 89.59 | 72.36 | 41.59 | 19.55 | Ox | 75.87 | 66.23 | 27.56 | 10.71 |
|  | Of | 84.98 | 69.99 | 41.59 | 16.99 | - | - | - | - | - |
| Angr | O0 | 91.44 | 98.29 | 22.13 | 80.00 | Od | 20.31 | 99.47 | 7.01 | 94.83 |
|  | O2 | 74.67 | 90.50 | 8.53 | 48.38 | O1 | 44.99 | 99.23 | 7.01 | 55.04 |
|  | O3 | 74.27 | 90.35 | 16.71 | 45.25 | O2 | 23.21 | 99.07 | 10.79 | 50.78 |
|  | Os | 87.30 | 97.18 | 12.45 | 70.80 | Ox | 21.05 | 99.11 | 6.94 | 50.77 |
|  | Of | 72.76 | 90.42 | 16.90 | 45.25 | - | - | - | - | - |
| Bap | O0 | 96.36 | 85.55 | 33.59 | 33.43 | Od | 88.24 | 67.14 | 7.97 | 30.14 |
|  | O2 | 78.62 | 57.64 | 3.30 | 24.47 | O1 | 87.18 | 59.03 | 65.54 | 29.43 |
|  | O3 | 77.40 | 57.22 | 3.06 | 24.30 | O2 | 90.68 | 57.85 | 74.56 | 29.94 |
|  | Os | 88.25 | 65.44 | 8.78 | 25.97 | Ox | 90.62 | 57.13 | 74.55 | 23.71 |
|  | Of | 77.63 | 58.11 | 4.75 | 24.33 | - | - | - | - | - |
| Radare2 | O0 | 98.71 | 97.01 | 80.01 | 60.02 | Od | 97.27 | 77.74 | 7.96 | 30.24 |
|  | O2 | 97.39 | 51.90 | 20.86 | 23.99 | O1 | 98.05 | 64.69 | 93.08 | 29.63 |
|  | O3 | 97.12 | 49.66 | 21.58 | 23.81 | O2 | 97.26 | 62.26 | 92.54 | 30.02 |
|  | Os | 98.39 | 71.86 | 34.64 | 25.34 | Ox | 97.45 | 61.64 | 93.08 | 24.53 |
|  | Of | 97.52 | 51.44 | 21.87 | 23.84 | - | - | - | - | - |
| IDA | O0 | 99.47 | 92.64 | 87.34 | 36.28 | Od | 99.10 | 90.45 | 95.49 | 61.48 |
|  | O2 | 98.50 | 71.50 | 53.25 | 29.74 | O1 | 98.04 | 83.61 | 2.35 | 3.83 |
|  | O3 | 98.43 | 72.49 | 51.41 | 29.61 | O2 | 98.85 | 83.97 | 95.52 | 61.47 |
|  | Os | 98.35 | 81.97 | 74.39 | 31.49 | Ox | 98.86 | 83.02 | 95.48 | 57.79 |
|  | Of | 98.40 | 78.05 | 39.93 | 29.64 | - | - | - | - | - |
| Ninja | O0 | 97.24 | 99.78 | 38.75 | 85.88 | Od | 96.09 | 99.07 | 59.66 | 94.25 |
|  | O2 | 93.83 | 94.55 | 32.68 | 69.33 | O1 | 96.29 | 98.02 | 2.20 | 3.83 |
|  | O3 | 93.85 | 95.20 | 34.65 | 68.82 | O2 | 96.00 | 98.45 | 63.50 | 82.55 |
|  | Os | 96.40 | 98.63 | 39.64 | 73.38 | Ox | 95.99 | 98.83 | 55.89 | 94.25 |
|  | Of | 93.74 | 95.30 | 36.66 | 68.83 | - | - | - | - | - |
| Nucleus | O0 | 97.41 | 98.29 | 53.58 | 80.00 | Od | 88.33 | 97.63 | 40.78 | 94.44 |
|  | O2 | 94.99 | 90.49 | 48.11 | 66.79 | O1 | 88.53 | 96.60 | 40.78 | 93.46 |
|  | O3 | 95.15 | 90.76 | 48.11 | 66.21 | O2 | 85.34 | 96.48 | 40.73 | 94.00 |
|  | Os | 96.80 | 91.13 | 49.86 | 69.20 | Ox | 85.53 | 96.79 | 40.78 | 94.26 |
|  | Of | 94.31 | 90.44 | 46.53 | 66.14 | - | - | - | - | - |
| ByteWeight | O0 | 99.73 | 99.35 | 97.33 | 97.99 | Od | 95.62 | 92.25 | 49.58 | 52.43 |
|  | O2 | 97.54 | 96.09 | 78.15 | 83.61 | O1 | 95.87 | 95.78 | 73.99 | 63.50 |
|  | O3 | 97.94 | 97.33 | 77.67 | 89.91 | O2 | 97.98 | 96.34 | 74.81 | 80.68 |
|  | Os | 97.83 | 96.86 | 85.38 | 76.83 | Ox | 97.99 | 95.33 | 74.23 | 72.32 |
|  | Of | 97.74 | 96.82 | 87.12 | 84.47 | - | - | - | - | - |

TABLE XI: Statistics of false positives in function detection. *Mismatch*, *J-Tab*, *Scan*, *T-Call*, *Mis-disa* respectively mean wrong pattern matching, wrong handling of jump tables, wrong detection of tail calls, and wrong disassembly.

| Tools | Percentage of False Positives (%) | | | | | |
|---|---|---|---|---|---|---|
|  | Mismatch | J-Tab | Scan | T-Call | Mis-disa | Other |
| Dyninst | 80.80 | 0.00 | 0.00 | 19.20 | 0.00 | 0.00 |
| Ghidra | 0.01 | 0.00 | 0.00 | 0.05 | 0.00 | 99.94 |
| Ghidra-NE | 28.38 | 0.00 | 0.00 | 71.61 | 0.01 | 0.00 |
| Angr | 9.89 | 0.00 | 78.41 | 11.70 | 0.00 | 0.00 |
| Angr-NS | 92.98 | 0.00 | 0.00 | 7.01 | 0.01 | 0.00 |
| Bap | 99.99 | 0.00 | 0.00 | 0.00 | 0.01 | 0.00 |
| Radare2 | 6.83 | 0.00 | 0.00 | 0.00 | 0.01 | 93.17 |

TABLE XII: Statistics of false negatives in function detection. *J-Tab*, *T-Call*, *Non-Ret*, and *FP-Overlap* respectively represent missing jump table targets, missing tail calls, missing non-returning functions, and side effects of other false positives.

| Tools | Percentage of False Negatives (%) | | | | |
|---|---|---|---|---|---|
|  | J-Tab | T-Call | Non-Ret | FP-Overlap | No-Match |
| Dyninst | 0.03 | 1.68 | 2.07 | 10.16 | 86.06 |
| Ghidra | 0.02 | 0.05 | 1.48 | 12.42 | 86.03 |
| Ghidra-NE | 0.01 | 0.18 | 14.16 | 0.84 | 84.81 |
| Angr | 0.08 | 13.05 | 8.25 | 73.61 | 5.01 |
| Angr-NS | 1.18 | 1.06 | 4.01 | 32.78 | 60.97 |
| Bap | 1.32 | 2.42 | 7.99 | 11.82 | 76.45 |
| Radare2 | 1.12 | 2.78 | 3.64 | 5.52 | 86.94 |

We also notice three other interesting observations. First, commercial tools do better than open source ones. In particular, BINARY NINJA can identify over 97% of the functions with a precision over 95%. Second, NUCLEUS achieves comparable coverage and precision to BINARY NINJA, showing a high promise of its CFG-connectivity based solution. Based on an official blog [84], BINARY NINJA incorporates NUCLEUS for CFG and function detection. Third, BYTEWEIGHT outperforms BAP despite BAP internally runs BYTEWEIGHT. This is because BAP uses pre-trained signatures but BYTEWEIGHT uses signatures trained with our benchmark binaries.

**Use of Heuristics:** In function entry identification, two major heuristics are used. The first heuristic searches function entries using common prologues/data-mining models. We summarize the contribution and accuracy of the heuristic in Table XIX in Appendix D. Without counting functions recursively reachable from the matched ones, this heuristic recovers 17.36% more functions with an average precision of 77.53%. Also observable is that utility of this heuristic varies across optimization levels and architectures. Moreover, existing tools use different patterns or data-mining models, competing for a coverage-accuracy trade-off. Comparing to GHIDRA-NE and RADARE2, ANGR-NS and DYNINST use more aggressive patterns/models, producing higher coverage (21.3%/24.02 *v.s.* 18.24%/7.93%) but lower precision (56.61%/85.37% *v.s.* 98.42%/87.29%).

The other heuristic, used by ANGR, takes the begin of each code region detected by linear scan as a function entry. The heuristic recovers 23% more functions but reduces the precision by 26.96% because it often considers the begin of padding or data-in-code as a function entry.

**Understanding of Errors:** Table XI and XII present the analysis of false positives and false negative (the analysis of GHIDRA is only done on binaries with exception information).

For false positives, there are three common causes. First, the signature-based detection wrongly matches function entries (FP ratio - DYNINST: 80.80%, GHIDRA-NE: 28.38%, ANGR-NS: 92.98%, BAP: 99.99%). Second, inaccurate tail call detection takes target of a regular jump as a function entry (FP ratio - DYNINST: 19.20%, GHIDRA-NE: 71.61, ANGR: 11.70%). Third, incorrect disassembly results in erroneous call instructions to incorrect target functions (FP ratio - GHIDRA-NE: 0.01%, ANGR-NS: 0.01%, RADARE2: 0.01%). Beyond the three causes, ANGR produces 78.41% of its false positives because it considers code discovered by its linear scan as a function entry; GHIDRA generates 99.94% of its false positives because exception information also carries pointers to middle of functions; RADARE2, aggressively inferring code pointers based on xrefs, brings 93.17% of its false positives.

False negatives also have a group of similar causes. First, tools can miss targets of jump tables and fail to identify functions called by the target code or their successors (FN ratio - ANGR-NS: 1.18%, BAP: 1.32%, RADARE2: 1.12%). Jump tables have smaller impacts to ANGR and DYNINST because ANGR uses linear scan to compensate jump tables and DYNINST has high coverage of jump tables. Second, tools cannot recognize many tail calls and hence, miss the functions indicated by their targets (FN ratio - ANGR: 13.05%, BAP: 2.42%, RADARE2: 2.78%). Third, missed non-returning functions prevent the detection of many function entries in cases like Listing 8 in Appendix G (FN ratio - GHIDRA-NE: 14.16%, ANGR: 8.25%, ANGR-NS: 4.01%, BAP: 7.99%). Fourth, wrongly identified functions can over-lap with true functions, making the true ones un-recognizable (FN ratio - DYNINST: 10.16%, GHIDRA: 12.42%, ANGR: 73.61%, ANGR-NS: 32.78%). All the other functions are missed because they are neither reached by recursive descent nor matched by patterns (FN ratio - ANGR-NS: 60.97%, GHIDRA: 86.03%, RADARE2: 86.94%).

*4) CFG Reconstruction :* In this part, we measure 5 targets: (1) intra-procedure edges between basic blocks; (2) call graphs for direct calls; (3) indirect jumps and indirect calls; (4) tails calls; (5) non-returning functions. For task (1), we exclude the edge between a call and the fall through code. For jump tables in task (3), we consider a case with no targets resolved as a false negative. All other cases are counted as false positives. For task (4), we exclude recursive calls and we count unique target functions instead of jumps.

**Overall Performance:** Table XIII and XIV show the results of CFG reconstruction on Linux binaries and Windows binaries.

First, the tools can recover most of the edges with high accuracy. DYNINST, GHIDRA, and ANGR find over 90% of the edges with an accuracy higher than 95%. Moreover, the recovery of edges highly correlates to the recovery of instructions: precision and recall in the two tasks are consistent. The results of call graphs are similar to, so we omit the details.

Second, tools have different capabilities of handling jump tables. On average, GHIDRA and DYNINST can resolve over 93% of the jump tables with an accuracy of around 90%. RADARE2 and ANGR have a similar coverage rate (around 75%) while ANGR has much higher accuracy (96.27% *v.s.* 90%). In comparison to open-source tools, commercial tools have both higher coverage (96.5% *v.s.* 84.8%) and accuracy (99% *v.s.* 92.96%).

TABLE XIII: Results of CFG reconstruction on Linux. **Edge**, **CG**, **T-Call**, **N-Ret**, and **J-Tab** respectively mean edges, call graphs, tail calls, non-returning functions, and jump tables.

| | L | Edge | | CG | | T-Call | | N-Ret | | J-Tab | |
|---|---|---|---|---|---|---|---|---|---|---|---|
| | | Pre | Rec | Pre | Rec | Pre | Rec | Pre | Rec | Pre | Rec |
| Dyninst | O0 | 97.28 | 97.10 | 99.99 | 99.01 | 81.91 | 71.15 | 100.0 | 86.89 | 99.97 | 98.82 |
| | O2 | 98.61 | 92.41 | 99.99 | 91.17 | 68.75 | 78.84 | 100.0 | 86.04 | 99.87 | 99.05 |
| | O3 | 98.88 | 91.58 | 99.99 | 90.34 | 59.64 | 71.12 | 100.0 | 86.20 | 99.81 | 99.10 |
| | Os | 98.93 | 90.14 | 99.99 | 86.92 | 67.74 | 68.79 | 100.0 | 86.33 | 99.89 | 98.50 |
| | Of | 98.87 | 91.33 | 99.99 | 90.26 | 58.93 | 68.60 | 100.0 | 85.08 | 99.93 | 98.75 |
| Ghidra | O0 | 99.63 | 97.41 | 99.99 | 96.66 | 90.98 | 79.42 | 100.0 | 50.19 | 98.49 | 75.58 |
| | O2 | 99.89 | 89.61 | 99.99 | 88.91 | 95.93 | 97.91 | 99.98 | 71.21 | 97.46 | 97.55 |
| | O3 | 99.89 | 89.30 | 99.99 | 88.59 | 92.75 | 96.10 | 100.0 | 67.15 | 91.44 | 97.71 |
| | Os | 99.88 | 90.12 | 99.99 | 88.88 | 99.01 | 99.02 | 100.0 | 59.74 | 99.03 | 91.53 |
| | Of | 99.90 | 89.47 | 99.99 | 88.37 | 92.65 | 94.74 | 100.0 | 69.59 | 91.70 | 97.81 |
| Angr | O0 | 98.41 | 98.13 | 99.99 | 99.93 | 34.09 | 57.80 | 97.62 | 81.25 | 99.80 | 87.34 |
| | O2 | 94.88 | 94.95 | 99.99 | 99.98 | 91.20 | 77.46 | 89.57 | 80.48 | 89.21 | 70.07 |
| | O3 | 95.23 | 95.16 | 99.99 | 99.98 | 87.46 | 80.97 | 90.48 | 80.72 | 89.12 | 73.42 |
| | Os | 95.53 | 96.44 | 99.99 | 99.95 | 84.37 | 96.05 | 93.04 | 83.05 | 98.62 | 81.07 |
| | Of | 95.12 | 95.10 | 99.99 | 99.95 | 87.34 | 79.13 | 88.98 | 78.28 | 90.23 | 73.22 |
| Bap | O0 | 95.31 | 76.12 | 99.99 | 85.00 | - | - | 100.0 | 75.61 | - | - |
| | O2 | 91.76 | 55.48 | 99.99 | 78.05 | - | - | 99.18 | 76.66 | - | - |
| | O3 | 92.00 | 55.97 | 99.99 | 77.69 | - | - | 99.12 | 71.40 | - | - |
| | Os | 94.03 | 72.00 | 99.99 | 80.42 | - | - | 100.0 | 79.37 | - | - |
| | Of | 91.95 | 56.54 | 99.99 | 77.84 | - | - | 99.24 | 68.66 | - | - |
| Radare2 | O0 | 98.47 | 83.29 | 99.99 | 92.67 | - | - | 83.45 | 83.51 | 55.74 | 43.14 |
| | O2 | 98.07 | 85.75 | 99.99 | 81.02 | - | - | 96.93 | 79.80 | 98.11 | 97.78 |
| | O3 | 98.23 | 83.09 | 99.99 | 76.96 | - | - | 96.85 | 79.25 | 98.52 | 98.11 |
| | Os | 98.67 | 80.01 | 99.99 | 86.99 | - | - | 96.63 | 79.56 | 82.53 | 76.24 |
| | Of | 98.42 | 86.97 | 99.99 | 77.79 | - | - | 97.03 | 76.25 | 98.51 | 97.26 |
| IDA | O0 | 99.38 | 97.39 | 99.99 | 97.60 | 93.05 | 90.82 | 100.0 | 89.18 | 99.95 | 99.99 |
| | O2 | 99.32 | 92.74 | 99.99 | 90.35 | 72.57 | 87.79 | 100.0 | 89.32 | 99.88 | 99.59 |
| | O3 | 99.36 | 92.88 | 99.99 | 89.89 | 93.65 | 92.74 | 100.0 | 89.11 | 99.89 | 99.71 |
| | Os | 99.34 | 95.78 | 99.99 | 95.12 | 75.62 | 85.83 | 99.98 | 90.67 | 99.61 | 99.92 |
| | Of | 99.33 | 93.99 | 99.99 | 92.07 | 73.62 | 81.82 | 100.0 | 87.86 | 99.68 | 99.75 |
| Ninja | O0 | 99.99 | 99.46 | 99.99 | 99.75 | 54.98 | 87.69 | 100.0 | 86.28 | 99.60 | 99.30 |
| | O2 | 99.32 | 98.27 | 99.99 | 99.27 | 83.14 | 94.85 | 100.0 | 86.78 | 98.91 | 95.81 |
| | O3 | 99.20 | 94.97 | 99.99 | 98.89 | 84.18 | 93.37 | 100.0 | 86.84 | 98.67 | 89.87 |
| | Os | 99.58 | 99.05 | 99.99 | 99.55 | 87.24 | 96.91 | 100.0 | 90.13 | 99.02 | 96.36 |
| | Of | 99.39 | 95.87 | 99.99 | 98.85 | 84.79 | 90.93 | 100.0 | 86.65 | 98.77 | 90.26 |

Besides jump tables, there are two more types of indirect jumps: handle-written assembly code (336 cases like [48]) and indirect tail calls (35,087 cases). For the first type, GHIDRA, by analysing constant propagation, resolves 96 cases but only report one target in each case. BINARY NINJA generates results for 120 cases, however, with incorrect targets.

Third, ANGR, GHIDRA, IDA PRO, and BINARY NINJA have (limited) supports of indirect calls. ANGR resolves 43 cases but only reports one target in each case. Our manual analysis verified the targets are correct, all following the pattern of [mov/lea CONST, reg; ...; call reg]. GHIDRA finds an incomplete set of targets for 88,078 indirect calls (only one target for 87,947 cases). IDA PRO reports results for 15,325 indirect calls (only one target for 15,267 cases). The other 58 cases follow the format of Listing 9 in Appendix G, which are fully solved by IDA PRO. BINARY NINJA reports targets for 3,410 indirect calls (1 target for 92 cases; 2 targets for 1,865 cases; 3 targets for 469 cases; 4 targets for 889 targets; 4+ targets for 95 cases).

Fourth, existing tools are not perfect with detecting tail calls. The recovery rate ranges from 71.7% (DYNINST) to 91.28% (BINARY NINJA); the precision varies from 67.39% (DYNINST) to 90.21% (GHIDRA). We observe that many tools (ANGR, GHIDRA, BINARY NINJA) has the lowest precision with optimization disabled. This is because less tail calls are emitted at lower optimization levels and a few false positives can result in a low precision.

TABLE XIV: Results of CFG reconstruction on Windows.

| | W | Edge | | CG | | T-Call | | N-Ret | | J-Tab | |
|---|---|---|---|---|---|---|---|---|---|---|---|
| | | Pre | Rec | Pre | Rec | Pre | Rec | Pre | Rec | Pre | Rec |
| Ghidra | Od | 96.66 | 87.09 | 99.88 | 90.84 | 88.50 | 80.92 | 100.0 | 11.23 | 70.84 | 95.46 |
| | O1 | 98.15 | 82.78 | 99.91 | 85.24 | 85.64 | 88.41 | 100.0 | 7.37 | 77.30 | 97.79 |
| | O2 | 96.58 | 83.80 | 99.86 | 87.61 | 83.03 | 85.65 | 99.44 | 7.56 | 72.89 | 94.62 |
| | Ox | 96.56 | 84.19 | 99.83 | 88.48 | 83.44 | 87.85 | 99.44 | 7.51 | 72.91 | 94.96 |
| Angr | Od | 95.33 | 96.86 | 99.87 | 99.99 | 55.33 | 80.21 | 93.14 | 16.40 | 99.65 | 71.22 |
| | O1 | 96.43 | 97.80 | 99.89 | 99.99 | 97.30 | 91.87 | 96.67 | 35.85 | 99.82 | 73.65 |
| | O2 | 95.72 | 96.13 | 99.88 | 99.98 | 97.48 | 92.52 | 97.04 | 41.41 | 99.99 | 59.71 |
| | Ox | 95.58 | 96.12 | 99.86 | 99.98 | 96.61 | 92.48 | 97.11 | 42.05 | 99.99 | 60.03 |
| Bap | Od | 96.05 | 70.89 | 99.29 | 73.92 | - | - | 0.00 | 0.00 | - | - |
| | O1 | 97.00 | 80.00 | 99.99 | 80.87 | - | - | 100.0 | 1.52 | - | - |
| | O2 | 96.43 | 70.22 | 99.99 | 72.59 | - | - | 100.0 | 6.64 | - | - |
| | Ox | 96.40 | 72.77 | 99.99 | 69.23 | - | - | 100.0 | 6.16 | - | - |
| Radare2 | Od | 97.74 | 85.12 | 99.11 | 86.83 | - | - | 100.0 | 6.30 | 96.24 | 84.29 |
| | O1 | 98.97 | 80.03 | 99.90 | 81.14 | - | - | 100.0 | 6.43 | 88.89 | 71.56 |
| | O2 | 98.07 | 79.59 | 99.91 | 80.79 | - | - | 100.0 | 6.56 | 95.29 | 64.49 |
| | Ox | 98.03 | 80.71 | 99.90 | 81.26 | - | - | 100.0 | 6.55 | 95.16 | 64.03 |
| IDA | Od | 96.98 | 95.77 | 99.87 | 98.00 | 96.25 | 77.06 | 100.0 | 53.89 | 99.43 | 100.0 |
| | O1 | 98.36 | 95.79 | 99.87 | 97.41 | 90.95 | 81.80 | 100.0 | 58.09 | 99.66 | 99.81 |
| | O2 | 97.12 | 93.92 | 99.85 | 97.07 | 88.18 | 83.34 | 100.0 | 63.38 | 99.91 | 99.85 |
| | Ox | 97.11 | 93.73 | 99.84 | 96.88 | 88.32 | 81.99 | 100.0 | 63.19 | 99.66 | 99.81 |
| Ninja | Od | 98.63 | 97.39 | 99.90 | 99.75 | 84.89 | 80.37 | 100.0 | 38.66 | 96.97 | 93.09 |
| | O1 | 99.20 | 98.60 | 99.80 | 99.83 | 95.31 | 91.89 | 100.0 | 47.30 | 99.61 | 90.22 |
| | O2 | 98.59 | 96.63 | 99.88 | 98.93 | 95.49 | 92.76 | 100.0 | 48.67 | 96.49 | 91.49 |
| | Ox | 98.59 | 96.82 | 99.86 | 99.10 | 95.61 | 92.83 | 100.0 | 48.43 | 96.50 | 91.55 |

TABLE XV: Statistics of false positives in CFG-edge recovery. *Non-Ret*, *Func*, *T-Call*, *Inst*, *No-split*, *Jump-Tab* respectively represent undetected non-returning functions, unrecognized functions, unidentified tail calls, wrong instructions, missing edges to middle of basic blocks, and wrong jump table targets.

| Tools | Percentage of False Positives (%) | | | | | |
|---|---|---|---|---|---|---|
| | Non-Ret | Func | T-Call | Inst | No-split | J-Tab |
| **Dyninst** | 23.09 | 12.25 | 23.28 | 4.95 | 35.24 | 1.19 |
| **Ghidra** | 8.76 | 1.88 | 13.00 | 0.11 | 75.13 | 1.12 |
| **Angr** | 16.69 | 38.42 | 10.82 | 4.73 | 27.02 | 2.32 |
| **Bap** | 37.95 | 22.30 | 1.34 | 0.34 | 38.06 | 0 |
| **Radare2** | 73.32 | 3.90 | 15.55 | 0.67 | 4.23 | 2.34 |

Finally, tools can detect non-returning functions with very high precision (nearly 100% for many tools). However, they have limited coverage, especially on Windows binaries. Based on our analysis, the root cause of the low coverage is incomplete collection of non-returning library functions.

**Use of Heuristics:** The recovery of edges and call graphs share heuristics in disassembly so we omit the details.

The resolving of jump tables uses three heuristics. First, RADARE2, GHIDRA, and DYNINST use patterns to detect jump tables, identifying 85.53%, 98.01%, and 99.56% of all jump tables. Second, DYNINST and ANGR restrict the scope of slicing in VSA, missing the index bound in many jump tables. By enlarging the scope of slicing to 500 assignments in DYNINST and 10 basic blocks in ANGR, the two tools find the index bound in 2.1% and 19.8% more jump tables. Third, GHIDRA and RADARE2 discard jump tables with index over 1,024 and 512, respectively missing 51 and 2,435 jump tables.

To detect tail calls, RADARE2 relies on the distance between jumps and their targets. It is hard to pick a proper threshold: in our benchmark binaries, the distance for regular jumps (min: 0; max: 0xb5867) and tail calls (min: 0; max: 0xb507c5) are largely overlapped. GHIDRA uses the heuristic that a tail call and its target spans multiple functions. This heuristic is general to capture tail calls, producing a high coverage of 91.29%. However, it cannot recognize regular jumps between two parts of non-continuous functions, producing over 70K false positives. ANGR and DYNINST use control-flow and stack-height based heuristics, detecting 4.24% and 6.99% more tail calls. However, the heuristics are neither sound nor complete, producing both false negatives and false positives.

**Understanding of Errors:** Table XV presents the analysis of false positives in edge recovery. The false negatives of edge recovery are mostly caused by non-recovered instructions (see § IV-B1). We omit the details. For call graphs, the false positives and false negatives are generally side effects of errors in disassembly. Details are also discussed in § IV-B1.

The errors related to jump tables are case specific and tool specific. We randomly sample 10% false negatives and 10% false positives from each tool and do manual analysis to understand the reasons. The false negatives vary across tools. RADARE2, GHIDRA, and DYNINST rely on patterns to detect jump tables, leading to 64.1%, 31.4%, and 37.72% of their false negatives. Further, RADARE2 incurs 35.9% of its false negatives because it discards jump tables with over 512 entries; GHIDRA produces 62.63%, 5.50%, and 0.47% of its false negatives, respectively because it does not consider sub instructions in VSA analysis, does not capture the correct restrictions to the index, and discards jump tables with more than 1,024 entries; ANGR generates false negatives because of no VSA results (67.86%), wrong VSA results leading to over 100,000 entries (14.29%), and no handling of sbb instructions (17.86%); DYNINST's remaining false negatives are caused by its restriction of slicing scope (2.1%) and no handling of get_pc_thunk (60.18%). Beyond the above reasons, we also observe 6 cases where the index is unrestricted (*e.g.*, Listing 10), leading to false negatives in many tools. The causes of false positives are also diverse. RADARE2's false positives are mostly because it matches the wrong cmp/sub instruction (54.29%) or the restriction to the index is not by cmp/sub (45.71%). DYNINST produces false positives because it does not handle special aliases of the index (78.96%), does not consider restrictions by type (17.95%), and restricts the scope of slicing (3.09%). ANGR incurs false positives due to incorrect VSA solving (78.26%) and ignoring certain paths in slicing (21.74%). With GHIDRA, the false positives are because of wrongly considering indirect jumps as indirect calls (55.32%), no consideration of sub instruction in VSA analysis (34.04%), identifying extra/wrong restrictions to the index (4.26%), and including the default case as an entry (6.38%).

TABLE XVI: Statistics of false negatives in detection of non-returning functions. *Lib*, *Cond-NonRet*, *Exit-Inst*, *Fake-Ret*, *Propa* respectively stand for unrecognized non-returning library functions, argument-dependent non-returning functions (*e.g.*, `error`), instructions that exit (*e.g.*, `ud2`), functions that contain `ret` instructions but do not actually return, and propagation of other types of false negatives.

| Tools | Percentage of False Negatives (%) | | | | |
|---|---|---|---|---|---|
| | *Lib* | *Cond-NonRet* | *Exit-Inst* | *Fake-Ret* | *Propa* |
| **Dyninst** | 12.30 | 43.86 | 0.30 | 27.64 | 15.90 |
| **Ghidra** | 3.97 | 13.01 | 0.04 | 9.54 | 73.44 |
| **Angr** | 19.08 | 21.70 | 0.00 | 22.29 | 36.93 |
| **Bap** | 24.45 | 22.73 | 0.10 | 28.90 | 23.82 |
| **Radare2** | 19.89 | 26.04 | 0.08 | 13.07 | 40.92 |

In detecting tail calls, false negatives are largely caused by excluding conditional jumps (FN ratio - GHIDRA: 21.6%; ANGR 17.5%), missing boundaries between adjacent functions (FN ratio - GHIDRA: 78.4%), excluding targets reached by both conditional jumps and unconditional jumps (FN ratio - ANGR: 81.6%), incorrect calculation of stack height (FN ratio - ANGR: 0.9%), unrecognized functions and no stack adjustment patterns before the jumps (FN ratio - DYNINST:100%). For false positives, the most common cause is wrong function entries (FP ratio - GHIDRA: 5.9%; ANGR 19.7%; DYNINST: 2.4%). Other reasons include non-continuous functions (FP ratio - GHIDRA: 94.1%; ANGR 0.3%; DYNINST: 0.6%), instructions follow the format of `add esp/rsp, CONST` but do not tear down the stack (FP ratio - DYNINST:97%), unchanged stack height before the jump (FP ratio - ANGR: nearly 90%). We also notice that ANGR's false positives contain a group of conditional jumps, which are caused by a defect in its code.

Table XVI shows the false negatives in detecting non-returning functions. A special category is functions that have `ret` instructions but do not return (*e.g.,* `_Unwind_Resume` in Glibc). Such functions alter the stack to use a self-prepared return address, avoiding returning to the parent. We observe no false positives with DYNINST. RADARE2 generates false positives because of un-handled jump tables (23.40%), wrong function boundaries (4.26%), and propagation of other false positives (72.34%). ANGR produces false positives due to un-handled jump tables/tail calls (29.73%), wrong function boundaries (54.05%), and propagation of other false positives (16.22%). GHIDRA incurs all its false positives due to heuristic ㉙. False positives by BAP are due to implementation issues.

*C. Threats to Completeness and Validity*

Despite our best efforts to carefully evaluate the public versions of the selected tools, there could still be threats to the completeness and validity of our results. Completeness wise, we may have missed reporting the results of some tools in certain tasks due to implementation issues of the tools (other than fundamental limitations). For instance, we could not report Angr's results of symbolization with Windows binaries because of two implementation issues in Angr's Reassembler module (one is reported at https://github.com/angr/angr/issues/1998; the other one is because Reassembler tries to parses the PLT section that does not exist in Windows binaries). Validity wise, implementation issues in our evaluation and analysis code may have slipped into the results, leading to inaccurate reports of the tools' performance. To mitigate this threat, we checked if the evaluation results match our understanding of the tools. In particular, we checked if the false positives and false negatives were explainable by the tool's strategies.

## V. FINDINGS

This study uncovered the following major findings.

**F-1:** *Complex constructs are common and heuristics are indispensable to handle them.* We observe a large amount of complex constructs in our benchmarks (Table XVII). In particular, we found 295 instances of data-in-code, complementary to previous research [5]. Heuristics are indispensable to handle the constructs: they are responsible for 49% of the instructions, 17% of the function entries, and most xrefs.

**F-2:** *Heuristics inherently introduce coverage-correctness trade-offs.* Heuristics significantly increase coverage, but concurrently induce new errors. As a counter-movement, new heuristics are devised to reduce the errors, leading to coverage-correctness trade-offs in nearly every disassembly phase.

**F-3:** *Tools complement each other.* As shown in Table III, existing tools use heuristics and algorithms that have overlaps but also many differences. This features tools different strengths, indicating that tool selection should be demand-specific.

**F-4:** *Broader and deeper evaluation is needed for future improvements.* We envision that the community may have insufficiently evaluated existing tools. This prevents comprehensive understanding about the limitations of the tools. We hope our study will bring a piece of basis to broaden and deepen the evaluation towards better binary disassembly.

## VI. CONCLUSION

We present a systematization of binary disassembly with a thorough study and comprehensive evaluation, centering around the perspective of algorithms and heuristics. Our study, via the comprehension on the source code of 9 disassembly tools, presents in-depth understanding of their strategies. Our evaluation separately measures the tools on each disassembly phase and on individual strategies, which fully unveils how much the heuristics are used, how much the heuristics contribute to disassembly, and what errors the heuristics can introduce. Throughout the study, we derive a group of new observations that can amend/complement previous understandings and also inspire future directions of binary disassembly.

ACKNOWLEDGMENTS

We would like to thank our shepherd Yan Shoshitaishvili and the anonymous reviewers for their feedback. This project was supported by the Office of Naval Research (Grant#: N00014-16-1-2261, N00014-17-1-2788, and N00014-17-1-2787). Any opinions, findings, and conclusions or recommendations expressed in this paper are those of the authors and do not necessarily reflect the views of the funding agency.

## APPENDIX

TABLE XVII: Statistics of complex constructs in our benchmark binaries.

| *Complex Constructs* | *Types* | *Cases / Prog / Bin* |
| --- | --- | --- |
| Data in code | Padding bytes | 6,445,649 / 236 / 3,788 |
|  | Hard-coded bytes | 295 / 3 / 18 |
|  | Jump tables | 21,586 / 57 / 426 |
| Indirect Jumps | Jump tables | 118,616 / 225 / 3,594 |
|  | Indirect tail-calls | 35,087 / 26 / 465 |
|  | Handle-written ones | 336 / 1 / 4 |
| Special functions | Overlapping functions | 2,050 / 6 / 49 |
|  | Multi-entry functions | 103 / 3 / 11 |
|  | Non-return functions | 29,668 / 228 / 3,712 |
| (Direct) Tail-calls | N/A | 503,527 / 236 / 3,781 |

## A. Complex Constructs

Table XVII presents the statistics of complex constructs from our benchmark binaries.

## B. Building Ground Truth for Windows Binaries

First, we follow a similar approach as described in [5] to gather instructions. We use the line table in debug information to find the first instruction of each source-code/assembly statement, followed by recursive disassembly (skipping indirect control transfers) to find instructions not covered by the line table. We also found that Visual Studio considers hard-coded bytes as code in the line table. We manually exclude such cases. Second, we rely on symbols to identify functions. The symbol of a function also carries a flag indicating whether it returns, aiding us to gather non-returning functions. Third, we enable the DEBUGTYPE:FIXUP option in Visual Studio such that all xrefs are preserved in the linking process. Finally, we identify jump tables based on xrefs (which contain the base address and entries of a jump table). For each xref, if its target location contains a list of other xrefs pointing to basic blocks in the same function, we consider the xref refers to a jump table and deem the list of xrefs at its target location as entries. For correctness, we further verify if the jump table entries correspond to `switch-cases` in the source code. Otherwise, we manually verify the correctness (since some jump tables are hand-crafted or compiled from `if-else` statements).

Our initial ground truth of non-returning functions (for both Linux and Windows binaries) can miss propagated cases. To this end, we expand our ground truth by running algorithms ⑭ and ⑮ with recursive updates.

## C. Configurations of Disassembly Tools

**OBJDUMP, PSI, UROBOROS:** We use the recommended options to run them. We parse their outputs for results.
**DYNINST:** We use its *ParseAPI* interface to perform recursive disassembly and we further enable *IdiomMatching* to include the decision-tree based function matching (which is excluded when we do testing without heuristics). We parse the returned structure of *ParseAPI* to get the disassembly results.
**ANGR:** We use its *CFGFast* interface for disassembly. When using this interface, we enable the *normalize* and *detect_tail_calls* arguments so that a basic block is split by a control transfer to the middle and tail calls are detected. When testing disassembly without heuristics, we disable the *force_complete_scan* and *function_prologues* arguments to *CFGFast*, preventing linear scan and function matching. To obtain the results, we interpret the CFG returned by *CFGFast* for results. We also excluded functions marked as "alignment". We use the *Reassembler* interface for symbolization.
**GHIDRA:** Besides default settings, we enable *Assume Contiguous Functions Only* and *Allow Conditional Jumps* to perform tail call detection. We disable the *X86 Constant Reference Analyzer* in symbolization as it introduces many dummy xrefs. Finally, to test GHIDRA without heuristics, we disable *Function Start Search* and xrefs related options, to prevent the signature-based matching and xref based disassembly.

TABLE XVIII: Results of disassembly without heuristics.

|  | L | Instructions | | | | W | Instructions | | | |
|---|---|---|---|---|---|---|---|---|---|---|
|  |  | Avg | | Min | | | Avg | | Min | |
|  |  | Pre | Rec | Pre | Rec | | Pre | Rec | Pre | Rec |
| Dyninst | O0 | 99.99 | 66.30 | 99.97 | 1.20 | - | - | - | - | - |
|  | O2 | 99.99 | 66.44 | 99.91 | 1.19 | - | - | - | - | - |
|  | O3 | 99.99 | 63.79 | 99.91 | 1.11 | - | - | - | - | - |
|  | Os | 99.99 | 68.58 | 99.96 | 1.31 | - | - | - | - | - |
|  | Of | 99.99 | 66.35 | 99.78 | 1.11 | - | - | - | - | - |
| Ghidra-NE | O0 | 99.99 | 64.53 | 99.90 | 1.22 | Od | 99.87 | 53.34 | 93.66 | 2.26 |
|  | O2 | 99.99 | 69.97 | 99.62 | 1.13 | O1 | 99.87 | 53.58 | 94.66 | 2.61 |
|  | O3 | 99.99 | 63.28 | 99.63 | 1.21 | O2 | 99.87 | 53.24 | 95.32 | 2.15 |
|  | Os | 99.99 | 68.53 | 99.77 | 1.35 | Ox | 99.86 | 52.91 | 93.56 | 2.15 |
|  | Of | 99.99 | 64.67 | 99.64 | 1.27 | - | - | - | - | - |
| Ghidra | O0 | 99.99 | 91.03 | 99.96 | 14.08 | Od | 99.87 | 53.34 | 93.66 | 2.26 |
|  | O2 | 99.99 | 91.75 | 99.68 | 1.80 | O1 | 99.87 | 53.58 | 94.66 | 2.61 |
|  | O3 | 99.99 | 90.18 | 99.72 | 1.95 | O2 | 99.87 | 53.24 | 95.32 | 2.15 |
|  | Os | 99.99 | 93.59 | 99.94 | 19.49 | Ox | 99.86 | 52.91 | 93.56 | 2.15 |
|  | Of | 99.99 | 88.93 | 99.73 | 21.16 | - | - | - | - | - |
| Angr-NS | O0 | 99.99 | 65.78 | 99.99 | 1.21 | Od | 99.99 | 47.21 | 99.99 | 2.16 |
|  | O2 | 99.99 | 58.13 | 99.74 | 0.61 | O1 | 99.99 | 51.38 | 99.99 | 2.51 |
|  | O3 | 99.99 | 58.62 | 99.94 | 0.54 | O2 | 99.99 | 46.32 | 99.99 | 2.06 |
|  | Os | 99.99 | 64.14 | 96.46 | 0.80 | Ox | 99.99 | 45.98 | 99.98 | 2.06 |
|  | Of | 99.99 | 60.14 | 97.25 | 0.55 | - | - | - | - | - |
| Radare2 | O0 | 99.99 | 6.37 | 99.99 | 0.02 | Od | 99.99 | 2.88 | 99.96 | 0.03 |
|  | O2 | 99.99 | 11.87 | 99.99 | 0.02 | O1 | 99.99 | 3.69 | 99.99 | 0.04 |
|  | O3 | 99.99 | 12.03 | 99.99 | 0.01 | O2 | 99.99 | 4.49 | 99.92 | 0.03 |
|  | Os | 99.86 | 7.29 | 83.70 | 0.01 | Ox | 99.99 | 4.14 | 99.83 | 0.03 |
|  | Of | 99.87 | 7.94 | 79.25 | 0.01 | - | - | - | - | - |

TABLE XIX: Results of function-matching. The baseline of Rec is the total number of true functions.

|  | L | Function Matching | | | | W | Function Matching | | | |
|---|---|---|---|---|---|---|---|---|---|---|
|  |  | Avg | | Min | | | Avg | | Min | |
|  |  | Pre | Rec | Pre | Rec | | Pre | Rec | Pre | Rec |
| Dyninst | O0 | 97.56 | 45.64 | 50.00 | 0.99 | - | - | - | - | - |
|  | O2 | 83.84 | 33.30 | 17.24 | 0.72 | - | - | - | - | - |
|  | O3 | 83.08 | 33.88 | 19.35 | 0.64 | - | - | - | - | - |
|  | Os | 89.11 | 4.88 | 0.00 | 0.00 | - | - | - | - | - |
|  | Of | 73.24 | 2.39 | 0.00 | 0.00 | - | - | - | - | - |
| Ghidra-NE | O0 | 97.26 | 42.43 | 0.00 | 0.00 | Od | 99.78 | 30.32 | 94.42 | 3.00 |
|  | O2 | 97.29 | 3.48 | 0.00 | 0.00 | O1 | 99.62 | 18.38 | 86.78 | 1.78 |
|  | O3 | 97.09 | 4.42 | 0.00 | 0.00 | O2 | 99.68 | 20.98 | 93.58 | 0.92 |
|  | Os | 97.69 | 16.55 | 0.00 | 0.00 | Ox | 99.69 | 23.08 | 93.58 | 0.89 |
|  | Of | 97.72 | 4.49 | 0.00 | 0.00 | - | - | - | - | - |
| Angr-NS | O0 | 91.04 | 36.93 | 0.00 | 0.00 | Od | 45.65 | 17.35 | 1.04 | 0.16 |
|  | O2 | 53.02 | 19.20 | 0.00 | 0.00 | O1 | 51.90 | 12.95 | 2.60 | 0.16 |
|  | O3 | 54.97 | 21.12 | 0.00 | 0.00 | O2 | 47.56 | 12.60 | 6.34 | 0.50 |
|  | Os | 59.15 | 26.16 | 0.00 | 0.00 | Ox | 47.93 | 13.36 | 6.86 | 0.50 |
|  | Of | 58.26 | 22.03 | 0.00 | 0.00 | - | - | - | - | - |
| Bap | O0 | 89.10 | 31.67 | 5.68 | 1.78 | Od | 57.94 | 12.78 | 1.16 | 0.23 |
|  | O2 | 54.21 | 16.23 | 0.00 | 0.00 | O1 | 50.45 | 7.57 | 1.11 | 0.23 |
|  | O3 | 55.03 | 18.21 | 0.00 | 0.00 | O2 | 55.54 | 8.97 | 1.28 | 0.25 |
|  | Os | 66.83 | 16.18 | 0.00 | 0.00 | Ox | 55.58 | 8.40 | 1.16 | 0.23 |
|  | Of | 54.75 | 17.91 | 1.41 | 1.03 | - | - | - | - | - |
| Radare2 | O0 | 99.86 | 30.66 | 66.66 | 0.00 | Od | 77.36 | 14.83 | 0.00 | 0.00 |
|  | O2 | 97.19 | 0.46 | 50.00 | 0.00 | O1 | 74.53 | 7.30 | 0.00 | 0.00 |
|  | O3 | 97.15 | 0.37 | 50.00 | 0.00 | O2 | 71.80 | 5.24 | 0.00 | 0.00 |
|  | Os | 97.86 | 6.30 | 50.00 | 0.00 | Ox | 71.80 | 5.58 | 0.00 | 0.00 |
|  | Of | 98.05 | 0.60 | 58.33 | 0.00 | - | - | - | - | - |

**RADARE2:** We run RADARE2 with four options: *aa* for default recursive disassembly, *aanr* for non-return function detection, *aac* and *aap* respectively enable xref based disassembly and signature-based function matching. With *aac* and

*aap* disabled, we can test RADARE2 without heuristics. We use APIs provided by RADARE2 to obtain the results.

**BAP:** We run BAP with *-dasm* and *-drcfg* to do disassembly and reconstruct the CFG. Note the plugin for *-drcfg* needs extra installation. We use the *with-no-return* [27] pass to detect non-returning function. We parse BAP's outputs for results.

**IDA PRO, BINARY NINJA:** We run the two tools with default settings and use their public APIs to get the results.

### D. Breakdown Analysis of Heuristics

Table XVIII and XIX respectively show the results of instruction recovery without heuristics and the results of pattern-based function matching.

### E. Overlap of False Positives and False Negatives

TABLE XX: Overlap of FP and FN in disassembly.

| Type | Number of Appearance (%) | | | | | | | |
|------|---|---|---|---|---|---|---|---|
| | 1 | 2 | 3 | 4 | 5 | 6 | 7 | 8 |
| FP | 92.05 | 4.11 | 2.52 | 0.85 | 0.31 | 0.14 | 0.01 | 0.001 |
| FN | 83.08 | 14.16 | 2.36 | 0.34 | 0.04 | 0.01 | 0.0005 | 0 |

TABLE XXI: Overlap of FP and FN in symbolization.

| Type | Number of Appearance (%) | | | | | |
|------|---|---|---|---|---|---|
| | 1 | 2 | 3 | 4 | 5 | 6 |
| FP | 95.30 | 4.70 | 0.03 | 0.001 | 0.00 | 0.00 |
| FN | 85.60 | 5.40 | 9.00 | 0.02 | 0.00 | 0.00 |

TABLE XXII: Overlap of FP and FN in function detection.

| Type | Number of Appearance (%) | | | | | | | | | |
|------|---|---|---|---|---|---|---|---|---|---|
| | 1 | 2 | 3 | 4 | 5 | 6 | 7 | 8 | 9 | 10 |
| FP | 70.4 | 19.2 | 2.5 | 8.6 | 1.5 | 0.2 | 0.05 | 0.007 | 0.01 | 0.007 |
| FN | 32.9 | 8.1 | 4.1 | 7.3 | 14.7 | 14.5 | 7.8 | 5.6 | 3.2 | 1.7 |

TABLE XXIII: Overlap of FP and FN in CFG recovery.

| Type | Number of Appearance (%) | | | | | | |
|------|---|---|---|---|---|---|---|
| | 1 | 2 | 3 | 4 | 5 | 6 | 7 |
| FP (EDGE) | 75.2 | 14.8 | 4.6 | 2.5 | 2.2 | 0.4 | 0.2 |
| FN (EDGE) | 43.0 | 22.3 | 14.3 | 10.1 | 4.8 | 3.7 | 1.7 |
| FP (CG) | 99.9 | 0.01 | 0.01 | 0 | 0.01 | 0 | 0 |
| FN (CG) | 54.4 | 20.5 | 14.8 | 7.9 | 2.1 | 0.2 | 0.001 |
| FP (T-Call) | 91.3 | 7.9 | 0.7 | 0.03 | 0.0001 | 0 | 0 |
| FN (T-Call) | 61.1 | 29.5 | 7.1 | 0.9 | 1.4 | 0 | 0 |
| FP (Non-Ret) | 98.0 | 2.0 | 0 | 0 | 0 | 0 | 0 |
| FN (Non-Ret) | 48.5 | 20.7 | 8.3 | 7.0 | 6.0 | 3.4 | 6.2 |
| FP (J-Tab) | 74.8 | 17.1 | 6.1 | 1.9 | 0 | 0 | 0 |
| FN (J-Tab) | 44.8 | 36.7 | 12.9 | 4.5 | 1.1 | 0.01 | 0 |

In Table XX, XXI, XXII, and XXIII, we present the overlap of false positives (FP) and false negatives (FN) in different tasks. Each cell indicates the percentage of FPs/FNs produced by the number of tools specified by ***Number of Appearance*** (*e.g.,* the value (2.36%) in the cell [FN, 3] in Table XX means 2.36% of the FNs are produced by 3 tools).

### F. Understanding of Commercial Tools

We attempted to infer how IDA PRO and BINARY NINJA operate, based on empirical experiments, blogs [49, 76, 84], documentations [1], and communications with the developers.

**Disassembly:** Both IDA PRO and BINARY NINJA perform recursive descent to recover instructions. We inferred this based on their correct results of handling Listing 2 and [84]. They also take other approaches to handling code gaps.

BINARY NINJA follows a heuristic as described in [84] to deal with code gaps. It linearly scans non-disassembled code regions and aggregates call targets. Once done, BINARY NINJA sorts all the targets based on the times of being referenced and (in order) hands them off to further recursive descent.

IDA PRO at least uses four strategies to handle code gaps: (1) it searches for common code sequences (*e.g.*, [push bp; mov bp, sp]). We inferred this from the kernel option of *mark typical code sequences as code* [1] and the test-case in Listing 11; (2) it considers addresses in the .eh_frame sections for recursive descent. This is inferred from the *enable EH analysis* kernel option, confirmed by comparing the results with and without the .eh_frame section; (3) it also performs recursive descent at the targets of d2c xrefs. This is inferred from the *create function if data xref data→code32 exists* kernel option, verified by comparing the results with and without certain xrefs; (4) it coagulates the remaining bytes in .text as code or data (the *make final analysis pass* kernel option).

**Symbolization:** It is in general hard to infer how exactly IDA PRO and BINARY NINJA do symbolization. We empirically learned the following strategies: (1) The two tools do not use heuristics ❽, ❿, and ⓬ (2) BINARY NINJA rarely considers d2c and d2d xrefs. (3) The majority of c2d xrefs identified by BINARY NINJA are constant address operands. (4) IDA PRO will seek to symbolize the data unit at the target of a c2d xref. It will further seek to symbolize the neighbors of the data unit at the target location.

**Function Entry Identification:** Both IDA PRO and BINARY NINJA consider the targets of direct/indirect calls as function entries. They further apply some other approaches.

BINARY NINJA adopts the idea from [6] to facilitate identification of function entries. It traverse the inter-procedural CFG and groups all the connected basic blocks as a function. Similar to many open source tools, BINARY NINJA also considers targets of tail calls as function entries.

IDA PRO leverages at least two other strategies to identify function entries: (1) it considers certain (but not all) addresses in the .eh_frame section as function entries. This is confirmed by comparing the results before and after removing the .eh_frame section; (2) it considers the targets of certain d2c xrefs as function entries. We inferred this by comparing the results before and after we intentionally destroy some d2c xrefs. However, thus far we are not fully aware of how IDA PRO selects .eh_frame items and the xrefs.

**Indirect Jumps:** Based on [84], BINARY NINJA implements VSA to handle jump tables. As discussed in § IV-B4, BINARY NINJA also resolves 120 hand-crafted indirect jumps, however, with wrong targets. It remains unclear how BINARY NINJA internally handles the 120 cases.

According to [1], IDA PRO [49] relies on patterns to detect jump tables. We further crafted test-cases (*e.g.*, Listing 12) to demonstrate that IDA PRO does not use VSA analysis.

**Indirect Calls:** We infer how BINARY NINJA and IDA PRO handle indirect calls by examining their results with our benchmark binaries (recall § IV-B3).

All the targets found by BINARY NINJA are constants propagated to the call sites. According to [76], this is achieved by path-sensitive data-flow analysis to calculate the ranges or disjoint sets of values (or VSA in general). Based on our further communication with the developers, the scope of analysis is intra-procedural.

IDA PRO found two types of targets. The first type is propagated from constants in the current function. The second type all follows the format in Listing 9. We envision that IDA PRO uses data-flow like constant propagation to handle indirect calls and applies patterns to find function tables.

**Tail Calls:** BINARY NINJA considers a jump as a tail call if the target is outside of the current function and the stack has a zero offset [84]. IDA PRO does not particularly handle tail calls, as clarified by their technical support team.

**Non-returning Functions:** Our analysis shows that IDA PRO and BINARY NINJA are detecting a similar group of non-returning functions as DYNINST. The difference is mainly caused by the recognition of non-returning library functions. Further, the three tools have comparable precision and recalls. This indicates that IDA PRO and BINARY NINJA are using similar recursive strategies to DYNINST.

## G. Interesting Cases and Test Cases

```
1 popfq                        1 popfq
2 .byte 0xf3,0xc3              2 repz retq
3 .size AES_cbc_encrypt        3 nop
4 .align 64                    4 nop
5 .LAES_Te:; data in code      5 (bad);disassembly error
6 .long 0xa56363c6             6 movslq -0x5b(%rbx),%esp
```

Listing 2: An example of data-in-code. This example comes from `Openssl`, where hand-crafted data are appended after code (*left*). Both OBJDUMP and ANGR incur errors in this case.

```
1 mov 0x6ab8a0, %esi           1 ; wrong decoding
2 mov %rbx, %rdi               2 640069: add [%rax],%eax
                               3 64006b: add [%rax],%cl
1 ; data;                      4 ...
2 0x6ab8a0: 69 00 64 00        5 640126: call unwind
3 ...                          6 ; non-return
```

Listing 3: A false postive of xref in `Xalan_base`. GHIDRA wrongly identifies the operand `0x6ab8a0` (line 1 *left-upper*) as a pointer and makes erroneous disassembly.

```
1 ; load base address          1 ; jump table
2 1b: lea 0x8a(%rip),%r8       2 ac: 84 cb ec ff
3 22: movzbl %cl,%ecx          3 a0: dc cb ec ff
4 25: movslq (%r8,%rcx,4),%rax 4 b4: 04 cc ec ff
5 29: add %r8,%rax             5 b8: 14 cc ec ff
6 2c: jmpq *%rax               6 ...
```

Listing 4: Jump table with 4-byte entries in 64-bit Gold Linker.

```
1 004286ab: add 0xfb49eb,%rax
2 ...
3 00fb49eb: undefined1 [11760811]; invalid address
```

Listing 5: A xref error in `zeusmp_base`. Operand `fb49eb` points to invalid data but GHIDRA symbolizes it.

```
1 mov $0xe,%ebx
2 movzwl 0x4e33be(%rbx),%eax; operand = 0x4e33be

1 0x4e33b9: 00 00 00 00; end of .fini
2 0x4e33be: (bad); invalid memory address
3 0x4e33c0: 04 e3 3b 00; begin of .rodata
```

Listing 6: A xref in `Busybox`. At line 2 (*upper* part), the constant operand points to an address (`0x4e33be`) in-between `.finit` section and `.rodata` section (*lower* part).

```
1 804f16c: 61 00 00 00 f7 c1 04 08; Unicode
```

Listing 7: Xref missed by ANGR in `addr2line`. At `0x804f16c`, ANGR detects a Unicode and jumps to `0x804f171`, missing the pointer at `0x804f170`.

```
1 callq 42ae30; non-returning
2 nopw %cs:0x0(%rax,%rax,1)
3 00 00 00; padding
4 cmpq $0x0,0x68(%rsi); start of a function
5 je 406f38
```

Listing 8: A missed function entry (line 4). The disassembler assumes code after line 1 falls through and includes code at line 4 to the preceding function.

```
1 app_main = {                 1 void run_app(int app){
2    test_main,                2    ...
3    acpid_main,               3    app_main[app](argc);
4    ...                       4    ...
5 }                            5 }
```

Listing 9: An indirect call that can be handled by IDA PRO. This indirect call uses a target from a function table.

```
1 switch(which){               1 ; no restriction on %al
2    case 't': ...             2 sub $0x64,%eax
3    ...                       3 movzbl %al,%eax
4    default:                  4 movslq (%r10,%rax,4),%rax
5       undefined();           5 add %r10,%rax
6 }                            6 jmpq *%rax
```

Listing 10: A jump table with unrestricted index in `dwp`. The default case in source code transfers to undefined behaviors, which is compiled into jump tables without index restriction.

```
1 pop %ebp             1 pop %ebp             1 pop %ebp
2 jmp 8048430          2 jmp 8048430          2 jmp 8048430
3 ;new function        3 ;new function        3 ;new function
4 push %ebp            4 push %eax            4 db 50h ; P
5 mov %ebp, %esp       5 mov %ebp, %esp       5 db 89h
```

Listing 11: Test-case to infer IDA PRO's disassembly. The code has a function (line 4, *left* part) that carries a common prologue but is never referenced. IDA PRO correctly disassembles the code. After altering the instruction at line 4 (*middle* part), IDA PRO considers the code as data (*right* part).

```
1 cmp $4, %rax
2 ja .Ldefault
3 ...
4 cmp $0, %rax
5 jle .Ldefault
6 ...
7 jmp *branch_tbl(,%rax,8)
```

Listing 12: Hand-crafted jump table with `rax` as the index. Tools with VSA analysis, like BINARY NINJA and DYNINST, figure out `rax` ranges from 1 to 4. IDA PRO wrongly considers `rax` ranges from 0 to 4.